\documentclass[aps,prd,showpacs,preprintnumbers,twelvepoint,floatfix]{revtex4}
\usepackage{epsfig}
\usepackage{latexsym,amsmath,amssymb,amstext}
\usepackage{float}
\usepackage{graphicx}
\usepackage{color}

\newcommand\bef{\begin{figure}}
\newcommand\eef[1]{\label{fg:#1}\end{figure}}
\newcommand\beq{\begin{equation}}
\newcommand\eeq[1]{\label{#1}\end{equation}}
\newcommand\beqa{\begin{eqnarray}}
\newcommand\eeqa[1]{\label{#1}\end{eqnarray}}
\newcommand\bet{\begin{table}}
\newcommand\eet[1]{\label{tb:#1}\end{table}}

\newcommand\fgn[1]{Figure \ref{fg:#1}}
\newcommand\eqn[1]{eq.\ (\ref{#1})}
\newcommand\scn[1]{Section \ref{sec:#1}}

\newcommand\ie{{\sl i.e.\/}}

\newcommand\etal{{\sl et al.\/}}

\newcommand\wls{\chi_{\scriptscriptstyle L}}
\newcommand\op[1]{{\mathcal O}_{#1}}
\newcommand\Tr{{\rm Tr}\/}

\newcommand{\B}{{\scriptscriptstyle B}}

\newcommand{\mub}{\mu_\B}

\begin{document}
\title{Quark number susceptibilities and equation of state\\
 at finite chemical potential in staggered QCD with $N_t=8$}
\author{Saumen\ \surname{Datta}}
\email{saumen@theory.tifr.res.in}
\affiliation{Department of Theoretical Physics, Tata Institute of Fundamental
         Research,\\ Homi Bhabha Road, Mumbai 400005, India.}
\author{Rajiv\ V.\ \surname{Gavai}}
\email{gavai@theory.tifr.res.in}
\affiliation{Department of Theoretical Physics, Tata Institute of Fundamental
         Research,\\ Homi Bhabha Road, Mumbai 400005, India.}
\author{Sourendu\ \surname{Gupta}}
\email{sgupta@theory.tifr.res.in}
\affiliation{Department of Theoretical Physics, Tata Institute of Fundamental
         Research,\\ Homi Bhabha Road, Mumbai 400005, India.}

\begin{abstract}
We report the measurement of quark number susceptibilities (QNS) and
their temperature dependence from simulations of QCD with two flavours of
light dynamical staggered quarks at finite temperature on $8\times32^3$
lattices.  From the radius of convergence of the Taylor expansion we
estimate the critical end point.  We use a Pad\'e approximant to resum
the series expansion and compute the equation of state at finite chemical
potential, namely the baryon number density and its contribution to the
pressure. We also report the isothermal compressibility of QCD matter
at finite baryon density.  Finally we explore the freezeout conditions
for a measure of fluctuations.  We examine some sources of systematic
and statistical errors in all of these measurements.
\end{abstract}

\preprint{TIFR/TH/16-51}
\maketitle

\section{Introduction}\label{sec:intro}
Quark number susceptibilities (QNS) \cite{milc,mumbai} have become important
objects of study in recent years. They are important ingredients in
the determination of the phase diagram of QCD \cite{nt4} as well as the
equation of state (EOS) of strongly interacting matter \cite{mumbai,pushan}.
They are also of interest in experimental studies of event-to-event
fluctuations of conserved quantities \cite{plb,stara,starb}. We have
earlier presented results with two flavours of light dynamical quarks
with lattice spacing of $1/(4T)$ ($N_t=4$) \cite{nt4,pushan} and $1/(6T)$
($N_t=6$) \cite{nt6}. In this paper we push closer to the continuum
limit with momentum cutoff of $8T$. Some preliminary results from our
current study were discussed in \cite{qm12,lat13}.

The pressure excess of strongly interacting matter at finite temperature,
$T$, and baryon chemical potential, $\mub$, over that at $\mub=0$ is
$\Delta P(\mub,T)$. We use the Maclaurin series expansion of $\Delta P$
in powers of $\mub$,
\beq
   \frac{\Delta P(\mub,T)}{T^4} = \sum_n \frac{\chi_B^n(T)}{T^{4-n}}
      \;\frac{z^n}{n!},\qquad{\rm where\ } z=\frac{\mub}T,
\eeq{macl}
and, due to CP symmetry, the series only has terms in even $n$, starting
from $n=2$. The coefficients are baryon number susceptibilities (BNS). We
shall often use the notation $\chi_B$ to mean $\chi_B^2$.

The Maclaurin expansion of \eqn{macl} also gives us an expansion for
the first derivative, \ie, the baryon number $n(\mub,T)$, and
the second derivative, \ie, $\chi_B(\mub,T)$. At the QCD critical
point, $\{\mub^E,T^E\}$, there is a critical divergence
\beq
   \frac{\chi_B(\mub,T^E)}{(T^E)^2} \propto \frac1{|\mub^2-(\mub^E)^2|^\psi},
\eeq{critd}
where $\psi$ is a critical index. A Widom scaling
argument was used in \cite{pushan} to show that $\psi=0.79$ for the Ising
universality class, in which the QCD critical point is expected to lie.
We have demonstrated earlier that with sufficient statistics one can
estimate both $z_E=\mub^E/T^E$ \cite{nt4,nt6,pushan} and $\psi$ \cite{pushan}
from lattice determinations of a small number of the baryon number
susceptibilities.

More detailed information comes from the
QNS. Since we work with two flavours of quarks, there can be two independent
chemical potentials, which can be chosen in many ways \cite{choice}. One
choice is to use them to get number densities of the two flavours of
quarks. If we call these $\mu_u$ and $\mu_d$, then the QNS are
\beq
   \chi_{\ell m} = \left.\frac{\partial^{\ell+m}P}{\partial\mu_u^\ell\partial\mu_d^m}
      \right|_{\mu_u=\mu_d=0}.
\eeq{defchi}
The order of the susceptibility is $\ell+m$. In our lattice computations the
two flavours are degenerate, so $\chi_{\ell m}=\chi_{m\ell}$, and we choose
this freedom to set $\ell\ge m$.  The BNS are combinations of the QNS. 
More details can be found in \cite{mumbai}, whose notation we follow.

The QNS have been used to test simplified models of QCD, such as PNJL
models \cite{pnjl}, effective models based on Schwinger-Dyson resummations
of weak-coupling expansions \cite{dyson}, and hadron resonance gas
models \cite{hrg}. They have also been proposed as diagnostics for the
presence of composites in the plasma state of QCD \cite{koch}. Currently
the most interesting use of the BNS is to compare with experiments
\cite{stara}. All such attempts make the assumption that heavy-ion experiments
see signals from thermalized matter whose temperature and chemical potentials
are then extracted by comparison of experimental data with predictions of
equilibrium statistical mechanics. A non-trivial statement about experiments
is likely to arise only when such treatments of quite different data are
compared. It was pointed out in \cite{cpod,plb} that a comparison
of the lattice predictions and experiments can give either a new way of
setting the lattice scale \cite{glmrx}, or the freezeout $T$ and $\mub$
for each collider energy \cite{lfo}.

In the next section we present the details of our simulations, and the
necessity of using large fermionic statistics. After that we present
results on the QNS and the $\{\mub^E,T^E\}$. In the fourth
section we report our results on the equation of state, $n$ and $\Delta P$,
as well as the isothermal bulk compressibility, $\kappa$. In the fifth
section we deal with measures of fluctuations and how they relate to
experiments. In this section we point out a substantial source of theory
systematic errors, and suggest how to take care of them.
In the final section we summarize our main results and point
out certain interesting implications.

\section{Simulations and statistics}\label{sec:method}
\bet
\begin{tabular}{|c|c|rr|c|r|}
\hline
$\beta$ & $ma$ & \multicolumn{2}{c|}{$T/T_c$} & $N$ & $N_v$ \\
\hline
5.48 &	0.0144 & 0.90 &(1) & $400+10\times{\bf 140}$ & 2000 \\
5.49 &	0.0139 & 0.93 &(1) & $15000+250\times{\bf 400}$ & 2000 \\
5.50 &	0.0136 & 0.94 &(1) & $15000+125\times{\bf 737}$ & 2000 \\
5.51 &	0.0133 & 0.96 &(1) & $15000+250\times{\bf 480}$ & 2000 \\
5.52 &	0.0129 & 0.98 &(1) & $15000+125\times{\bf 684}$ & 2000 \\
5.53 &	0.0127 & 1.00 &    & $15000+250\times{\bf 377}$ & 2000 \\
5.54 &	0.0125 & 1.02 &(1) & $15000+250\times{\bf 375}$ & 2000 \\
5.60 &	0.0113 & 1.14 &(1) & $15000+250\times{\bf 100}$ &  800 \\
5.77 &	0.0083 & 1.53 &(2) & $15000+250\times{\bf 100}$ &  800 \\
5.96 &	0.00625& 2.07 &(4) & $15000+250\times{\bf 100}$ & 1600 \\
\hline
\end{tabular}
\caption{The details of the measurements on $8\times32^3$ lattices. The
 statistics of gauge field configurations ($N$) is reported as the number
 of MD trajectories discarded for thermalization plus the separation between
 configurations times the number of configurations used. Each MD trajectory
 was taken to be 6 MD time units long.}
\eet{run8}

\bef
\begin{center}
\includegraphics[scale=0.7]{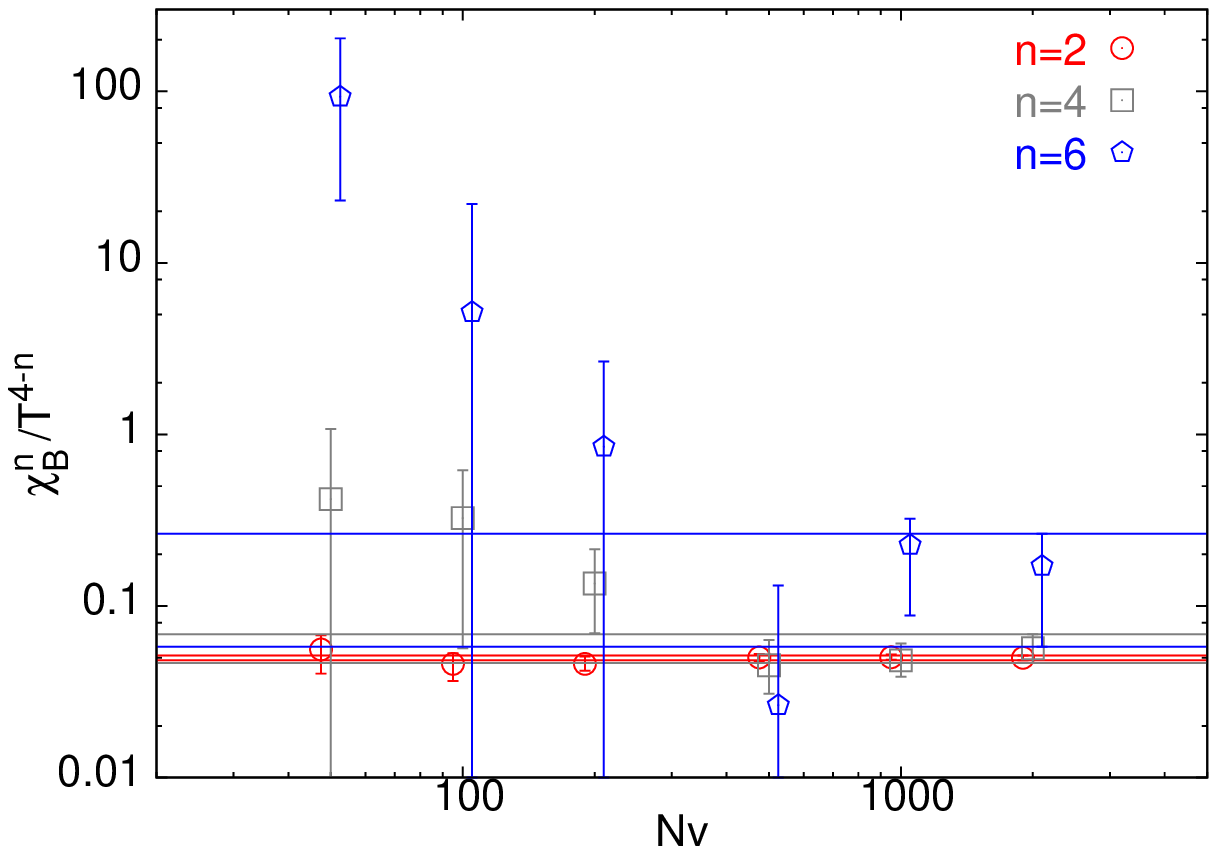}
\includegraphics[scale=0.7]{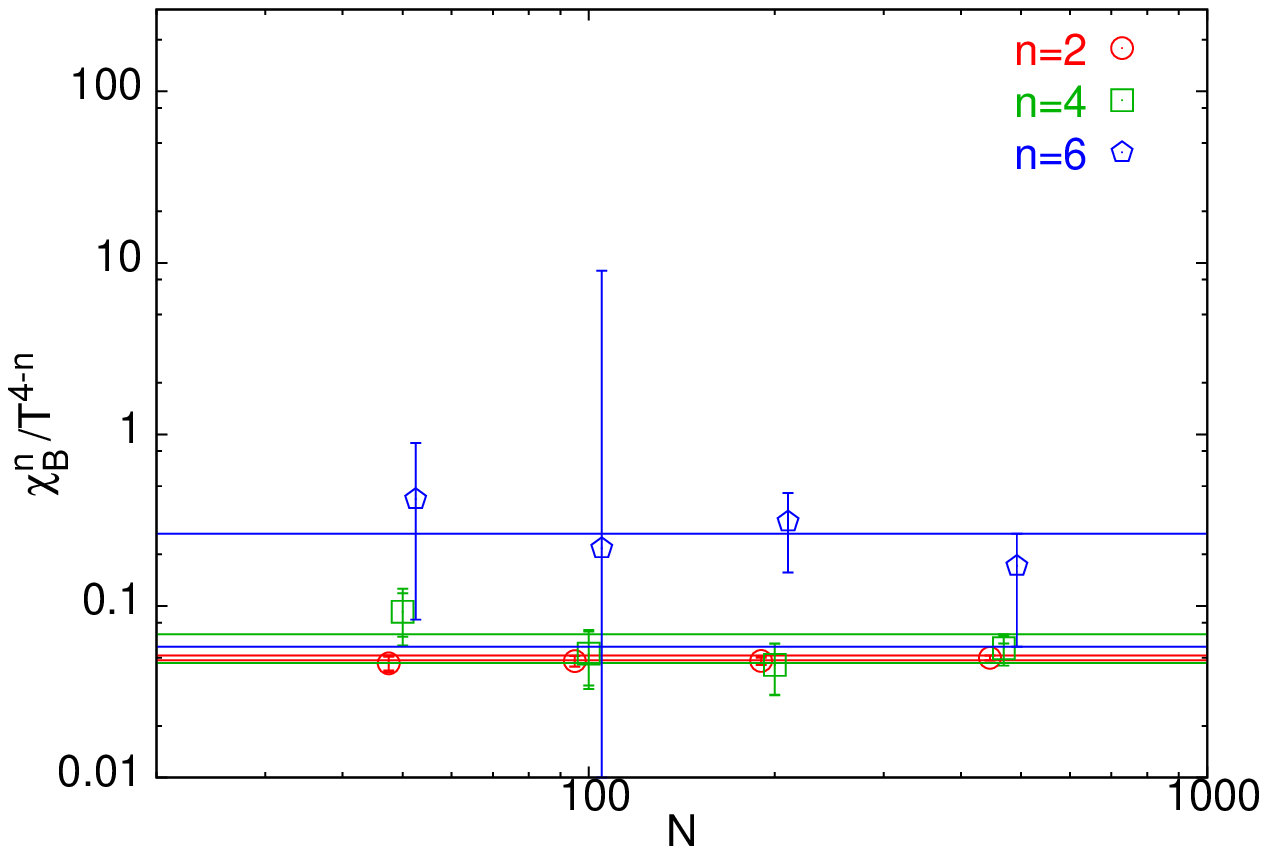}
\end{center}
\caption{The series coefficients of $\chi_{20}$ computed at $T/T_c=0.94\;
 (1)$, shown as functions of the number of fermion source vectors,
 $N_v$, (first panel) and the number of gauge configurations, $N$,
 (second panel) used. Note that $\chi^4$ is displaced slightly to the
 right in order to alleviate clutter.}
\eef{syst}

\bef
\begin{center}
\includegraphics[scale=0.45]{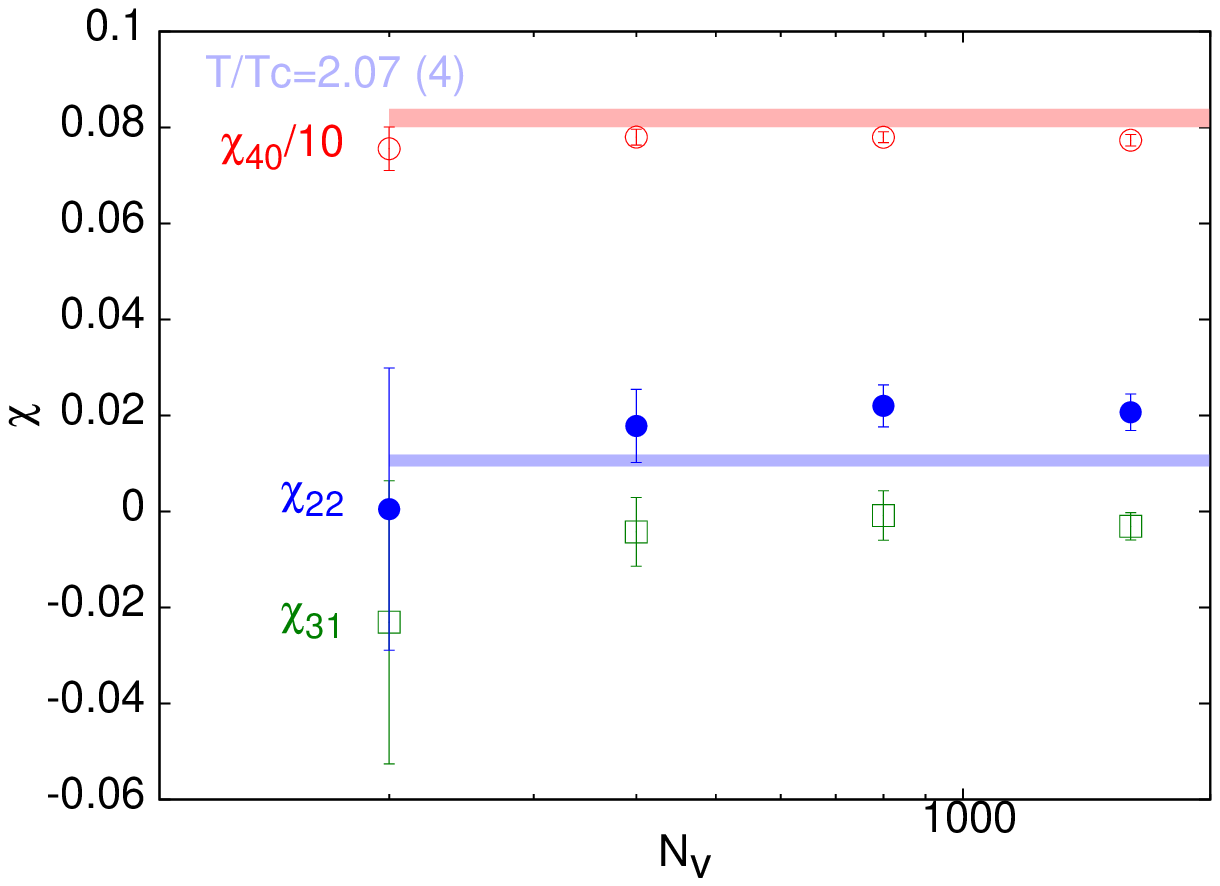}
\includegraphics[scale=0.45]{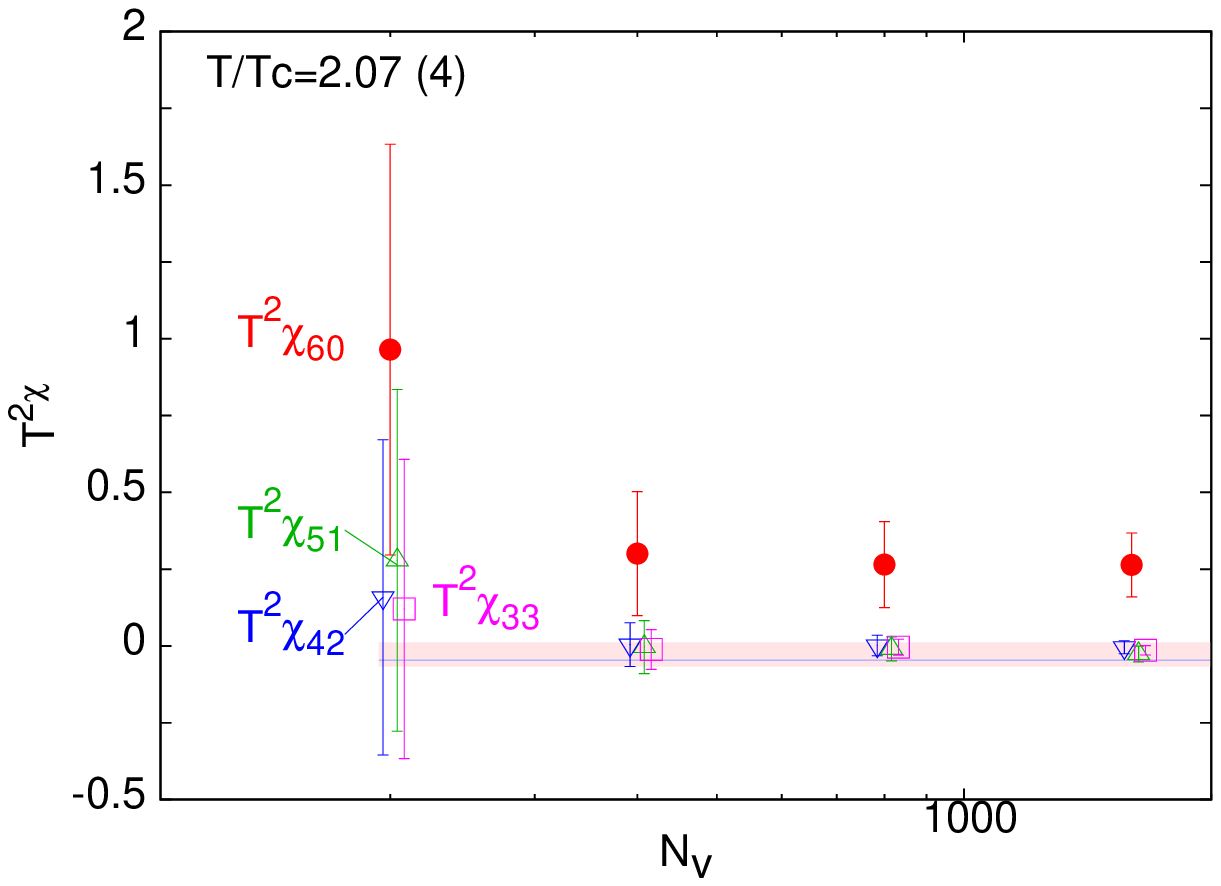}
\includegraphics[scale=0.45]{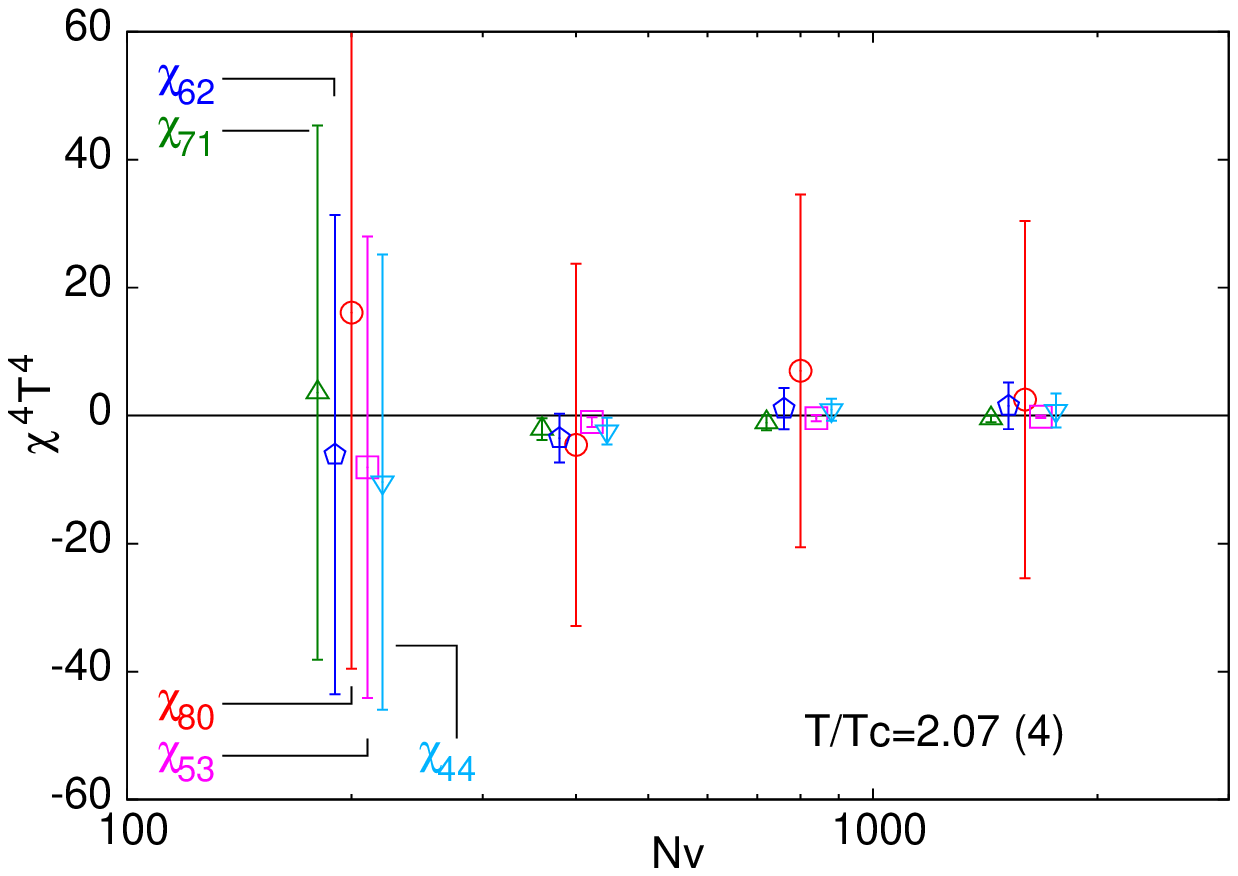}
\end{center}
\caption{The 4th order (first panel), 6th order (second panel) and the
 8th order (third panel) QNS at $2T_c$ as functions of the number of
 fermion source vectors, $N_v$, used. Different QNS for the same order
 have been displaced from each other for visibility. $\chi_{40}$ has been
 scaled down by a factor of 10 in order to fit into the scale shown.
 In the first panel the pink band is the DR prediction for $\chi_{40}$
 and the blue band for $\chi_{22}$ \cite{dr}.  In the second panel the
 pink band is the DR prediction for $\chi_{60}$ and the blue band for
 $\chi_{42}$ \cite{dr}.}
\eef{vecdephi}

Our simulations were carried out using two flavours of dynamical staggered
quarks with $m_\pi/m_\rho\simeq0.4$ and $N_t=8$. It has been known for
long that $N_f=2$ QCD at finite quark mass has no phase transition
\cite{rob}.  In \cite{wup} it was shown that there is no privileged
operator which measures a cross-over temperature.  In fact, a follow-up
study \cite{wup09} showed that the crossover in QCD with physical quark
masses is so broad that different measures of the crossover temperature
built using the renormalized chiral condensate alone gave results which
could differ by more than 7\% of their mean. This is about 5--6 times
the error bar on each. The difference between the crossover temperature
determined using the Wilson line and the chiral condensate is less than
3 times the error bar. The lesson for subsequent lattice studies is that
it is sufficient to use the simplest of measures of the crossover, unless
the goal is to improve the precision of the temperature scale.  We have
used the Wilson line susceptibility, as we had in earlier studies.
As a result, our temperature scales and other results are directly
comparable to our older results on coarser lattice spacings.

A previous study with staggered quarks on $N_t=8$ lattices \cite{milc93}
used the chiral condensate and the Polyakov loop, $L$, to locate a
crossover coupling in the range $5.52\le\beta\le5.56$.  We had earlier set
the scale for a study with coarser lattices ($N_t=6$) using the maximum
of the Polyakov loop susceptibility, $\wls$.  Two-loop scaling using a
non-perturbative value of the gauge coupling \cite{nt6} indicated that
the cross over measured by $\wls$ would lie close to $\beta=5.54$. In our
direct computation we found that the cross-over occurs at $\beta=5.53$.
Such a shift corresponds to an uncertainty in the temperature scale
of about a percent, and is expected when using two-loop scaling with
these cutoffs \cite{saumen}.  Using the new, finer, lattice cutoff,
we set the relative temperature scale as before. The systematic errors
in scale setting are estimated by comparing computations in different
renormalization schemes. We found that this error is about 1\%.

Numerical computation of operators with multiple fermion loops is time
consuming. Fast and accurate computations of the trace of a large and
sparse matrix, $A$, involves a noisy estimator: $2\Tr A= \overline{r^\dag
Ar}$ where $r$ is a complex vector drawn randomly from an appropriate
ensemble (here, Gaussian), over which we average. Every fermion loop is
such a trace over the lattice discretized Dirac operator. To get high
accuracy in products of traces, we need a large number, $N_v$, of random
fermion source vectors. Since the evaluation of thermal expectation
values of traces involves a Monte Carlo (over fermion sources) within a
Monte Carlo (over gauge configurations), we use a boot-strap over both
in order to estimate means and errors.

To stabilize measurements of the QNS, we require successively larger $N_v$
as the order increases. This is largely due to the increasing number
of fermion-line disconnected loops which can contribute as the order
increases. This can be seen even at the lowest non-trivial order, \ie,
for $\chi_{20}$. Below $T_c$, the operator $\op{11}$ (see \cite{nt4}
for a definition of these operators) contributes around 15\% of the
mean value, but about 50\% of the error. The situation is much worse
at $2T_c$ where this operator contributes about $0.2$\% of the mean,
but about 33\% of the error.  Fermion-line disconnected operators,
$\op{{n_1}{n_2}{n_3}\cdots}$, are the main source of noise since they
turn out to be fat-tailed \cite{pushan}.  At present there is no better
way of controlling them than by increasing $N_v$.

In \fgn{syst} we show that at low temperatures a reasonable estimate
of $\chi_B^n$ for $n=6$ requires $N_v$ as large as 1000. The number of
gauge configurations used is also important, and we have used a minimum
of around 400 gauge configurations at low temperatures in order to get
about 20\% errors for $n=6$.  As one sees from the data shown in the
figure, such statistics are essential.

At high temperature the problem simplifies a little. In \fgn{vecdephi}
we show how the estimates for the QNS depend on $N_v$ at $2T_c$. In the
high temperature phase we see that $N_v\simeq400$ is enough to control
errors.  Configurations are also smoother at high temperature, so the
number of gauge configurations required is also smaller.  $N_v\simeq400$
also begins to control the approach to an accurate measurement of 8th
order susceptibilities.  Of these the diagonal susceptibility $\chi_{80}
T^4$ has the largest errors; we can trace this to fluctuations in a
single noisy operator, $\op{2222}$. Since the
higher order QNS are divided by large factorials in Taylor expansions,
this degree of control over errors suffices for extrapolations to finite
chemical potential. The figure also demonstrates that the off-diagonal QNS
are in pretty good agreement with weak coupling computations at $2T_c$.

\section{Susceptibilities and the critical point}\label{sec:qns}
\bef
\begin{center}
\includegraphics[scale=0.7]{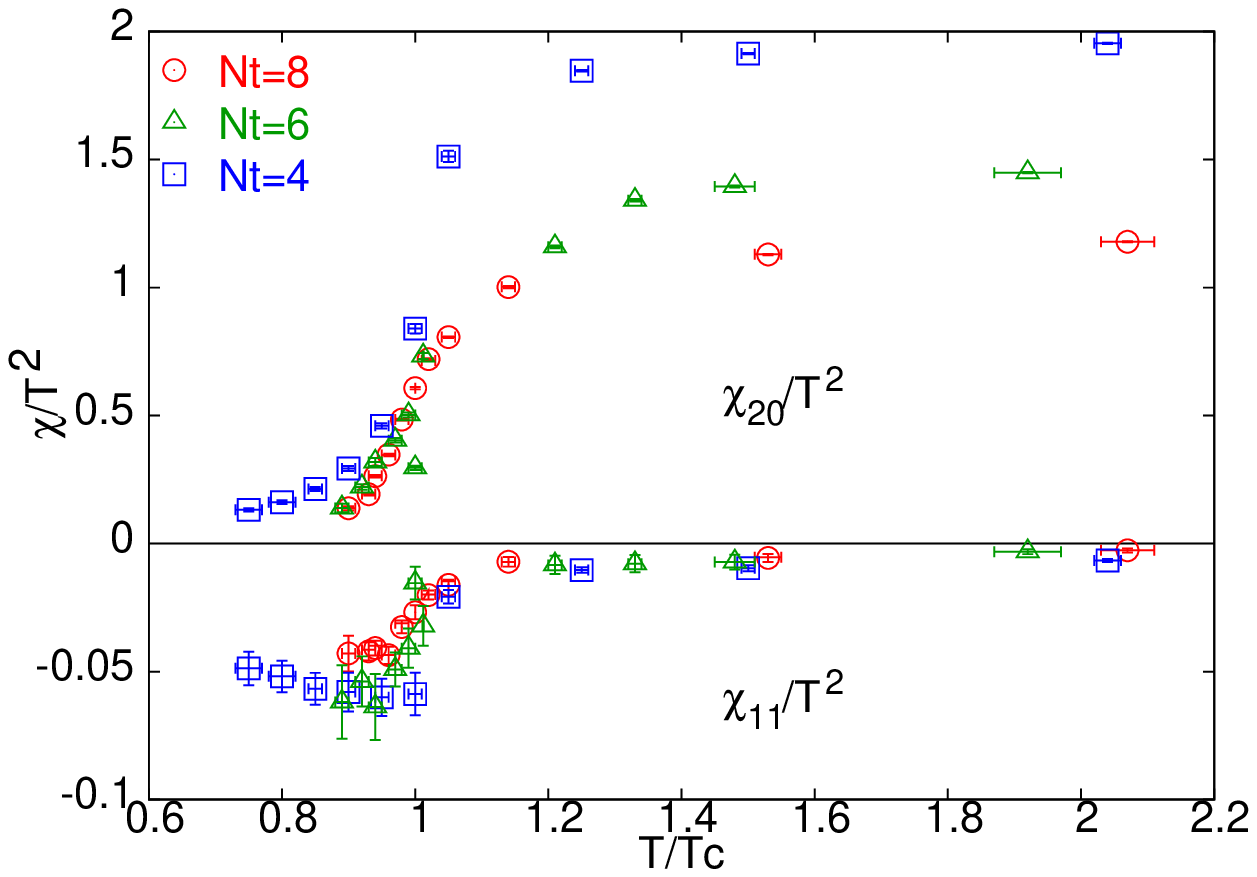}
\includegraphics[scale=0.7]{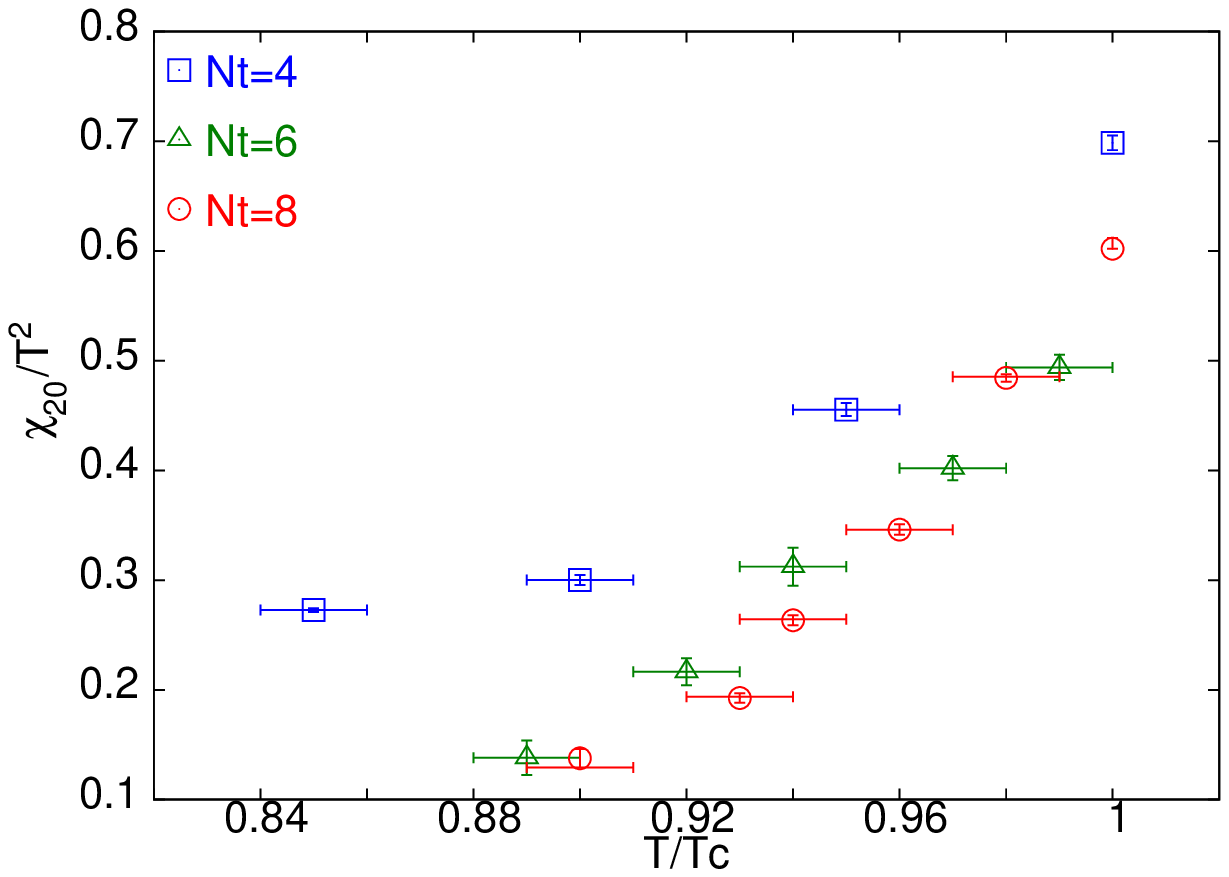}
\end{center}
\caption{The first panel compares $\chi_{20}/T^2$ and $\chi_{11}/T^2$ as a
  function of $T/T_c$ with different lattice cutoffs $a=1/(T N_t)$. Notice
  the difference in scale for the two quantities. In the second panel we
  zoom into the region of $T<T_c$, for $\chi_{20}/T^2$, since it is relevant
  to the position of the critical point, as we discuss later.}
\eef{ntdep}

In \fgn{ntdep} we show the second order QNS as functions of $T/T_c$ for
different lattice spacings. The measurements with lattice spacing $1/(4T)$
were presented in \cite{nt4} and recently updated in \cite{pushan}. The
data shown in \fgn{ntdep} for this cutoff comes from \cite{pushan}. For
lattice spacing of $1/(6T)$ we use the data of \cite{nt6}. Two
regions of temperature are clearly visible.  
Below $T_c$ the diagonal QNS, $\chi_{20}/T^2$, changes little
when the lattice cutoff is changed from $6T$ to $8T$. This is shown
more clearly in the zoom presented in the second panel. Below $T_c$
the data on $\chi_{11}/T^2$ from lattices with cutoff of $6T$ are seen
to be more noisy. This reflects the fact that these older measurements
used significantly smaller number of gauge configurations. This indicates
that finite lattice spacing effects are small for $T<T_c$ when going from $N_t=6$ to 8.

\bef
\begin{center}
\includegraphics[scale=0.65]{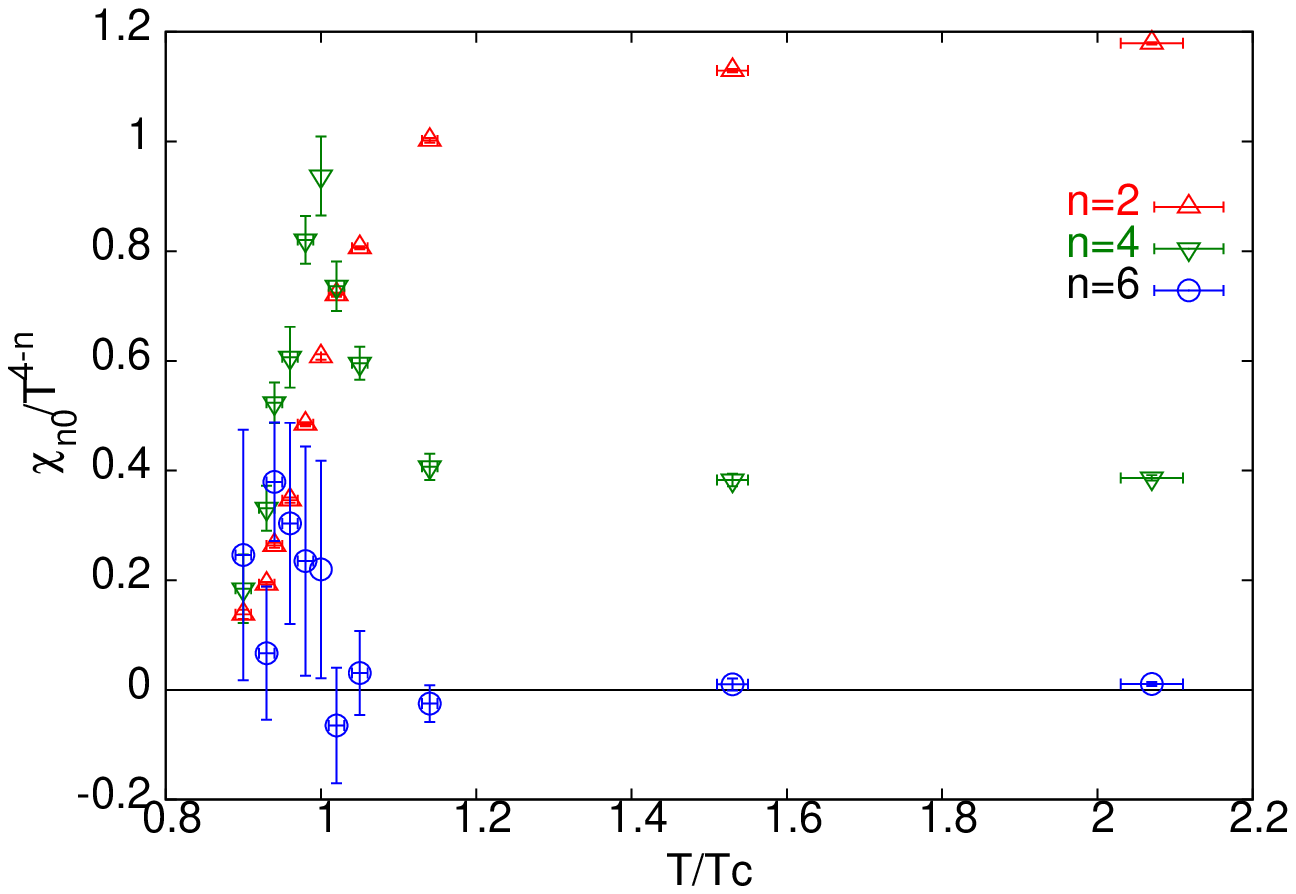}
\includegraphics[scale=0.65]{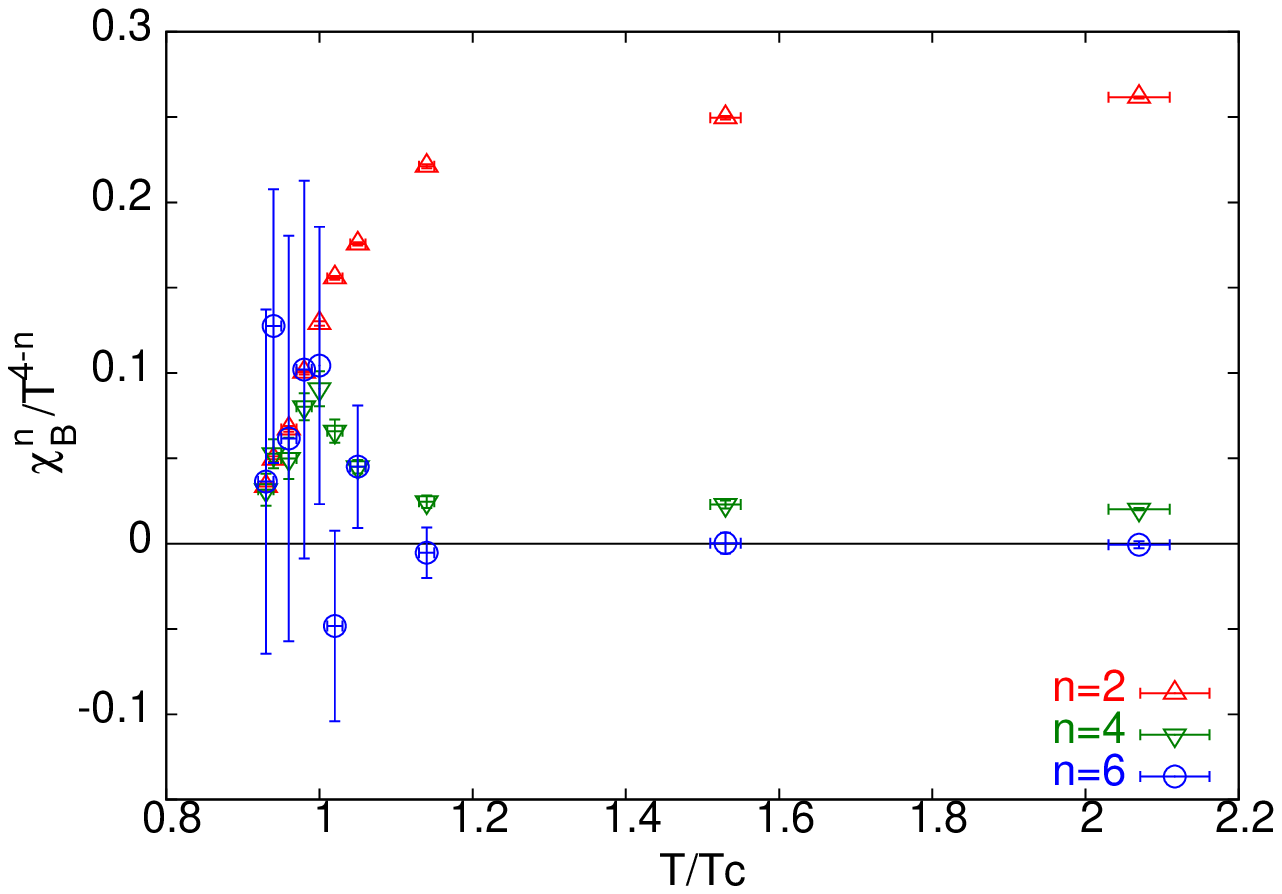}
\end{center}
\caption{Measurements of various susceptibilities for $N_t=8$. The
  diagonal QNS $\chi_{20}/T^2$ (up triangles), $\chi_{40}/2$ (down
  triangle) and $T^2\chi_{60}/24$ (circles) are shown in the first
  panel. In the second panel we show the baryon number susceptibilities,
  $\chi_B^2/T^2$ (up triangles), $\chi_B^4$ (down triangles) and
  $T^2\chi_B^6$ (circles).  }
\eef{series}

When the system is heated above $T_c$, 
$\chi_{11}/T^2$ is very close to zero in the weak-coupling
expansion, and that is also seen in this figure for $T>T_c$. However,
$\chi_{20}/T^2$ is seen to have a strong dependence on the lattice
spacing in this region of $T$.
The difference in the behaviour of the two second order QNS at high
temperature is due to an interesting phenomenon. There is an operator,
$\op2$,
which contributes to $\chi_{20}/T^2$
but not to $\chi_{11}/T^2$. This operator has a non-zero value for
free fermions, but is subject to a large finite lattice spacing effect.
Weak-coupling corrections to this operator are small. As a result, this
QNS shows strong finite lattice spacing effects. This was also seen
when the continuum limit of the quenched theory was taken \cite{qcont}.
Other QNS, including those shown in \fgn{vecdephi}, are reasonably close
to their continuum weak-coupling limit already for $N_t=8$.

In \fgn{series} we show two different series of susceptibilities.
The first panel shows diagonal QNS, $\chi_{n0}$.  A little below $T_c$
their magnitudes rise very rapidly with the order. A little above
$T_c$ the higher order QNS also approach their weak-coupling values.
$\chi_{40}$ peaks near $T_c$, and then approaches a non-vanishing
ideal-gas value for large $T$. $\chi_{60}T^2$ also seems to peak around
the same temperature, but it approaches zero above $T_c$.  In the second
part of this figure we show the BNS.  Their magnitudes also rise very
fast with order just below $T_c$.  While the second order QNS vary
monotonically with temperature, the fourth and sixth orders peak around
$T_c$ \cite{pushan}.  Since these QNS are measured using the same gauge
configurations and fermion source vectors, their errors are strongly
correlated. This cannot be shown in the figure, but is important for the
error estimation in all the succeeding analysis. Our bootstrap process is
designed to take care of these covariances.

\bef
\begin{center}
\includegraphics[scale=1.0]{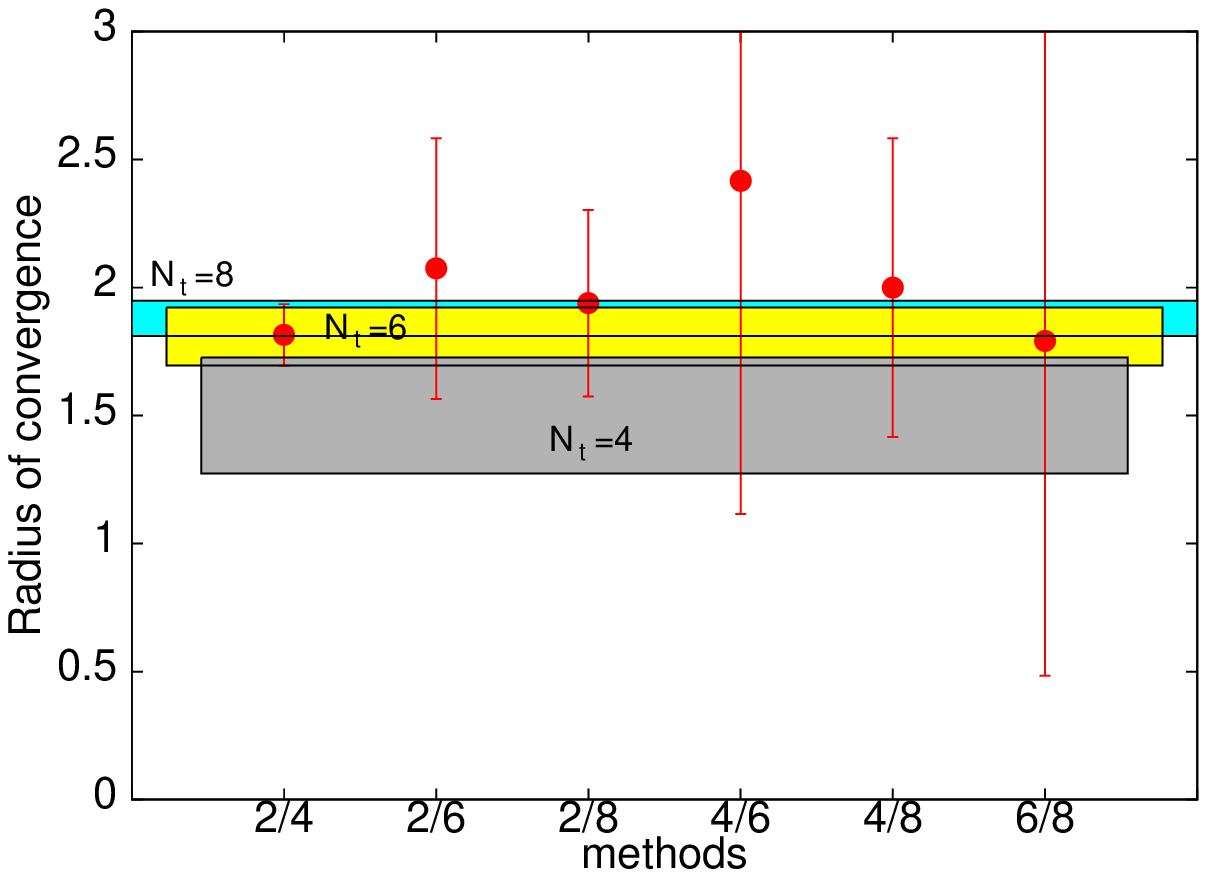}
\end{center}
\caption{Estimating the radius of convergence of the series at the
 temperature where it is found to be real. All the different estimators
 $\mub^{m/n}$ are shown (circles) along with the bootstrap estimator
 of their mean (the band marked $N_t=8$). Also shown are the bands of
 estimates obtained in previous computations.}
\eef{endpoint}

By comparing the coefficients of any two terms in the Maclaurin series
for $\chi_B$ induced by the expansion given in
\eqn{macl}, we have estimators of the radius of convergence
\beq
   \mu_B^{m/n} = \left[\frac{(n-2)!\chi_B^m}{(m-2)!\chi_B^n}\right]^{1/(n-m)}.
\eeq{estmn}
The radius of convergence generally corresponds to a singularity
at complex $\mu$. However, when all the $\chi^n$ are positive, then
this singularity lies on the real axis. In a bootstrap analysis, one
requires this criterion for all samples, since the mean would have an
imaginary part otherwise. We find that this selects out $\beta=5.50$,
which corresponds to a temperature of $0.94T_c$. In each bootstrap
sample one obtains all possible $\mu_B^{m/n}$.  These bootstrap
estimators are shown in \fgn{endpoint}. In each
bootstrap sample one may take an average over all the estimators. The
figure also shows the bootstrap estimator of the average, drawn as the
band labeled $N_t=8$. This gives the location of the critical point of QCD as
\beq
   \frac{\mub^E}{T^E} = 1.85\pm0.04, \qquad{\rm and}\qquad
   \frac{T^E}{T_c} = 0.94\pm0.01.
\eeq{cep}
This should be compared with the estimated $\mub^E/T^E=1.8\pm0.1$ for $N_t=6$
\cite{nt6} and the recent high-statistics determination $\mub^E/T^E=1.5\pm0.2$
for $N_t=4$ \cite{pushan}, which are also shown as bands labeled by $N_t$
in \fgn{endpoint}. Note again that our convention is to choose $T_c$ to be
the temperature at which the Polyakov loop susceptibility peaks. The
continuum value for this quantity was reported in \cite{tcpcont}

\section{Critical behaviour and equation of state}\label{sec:eos}
\bef
\begin{center}
\includegraphics[scale=1.0]{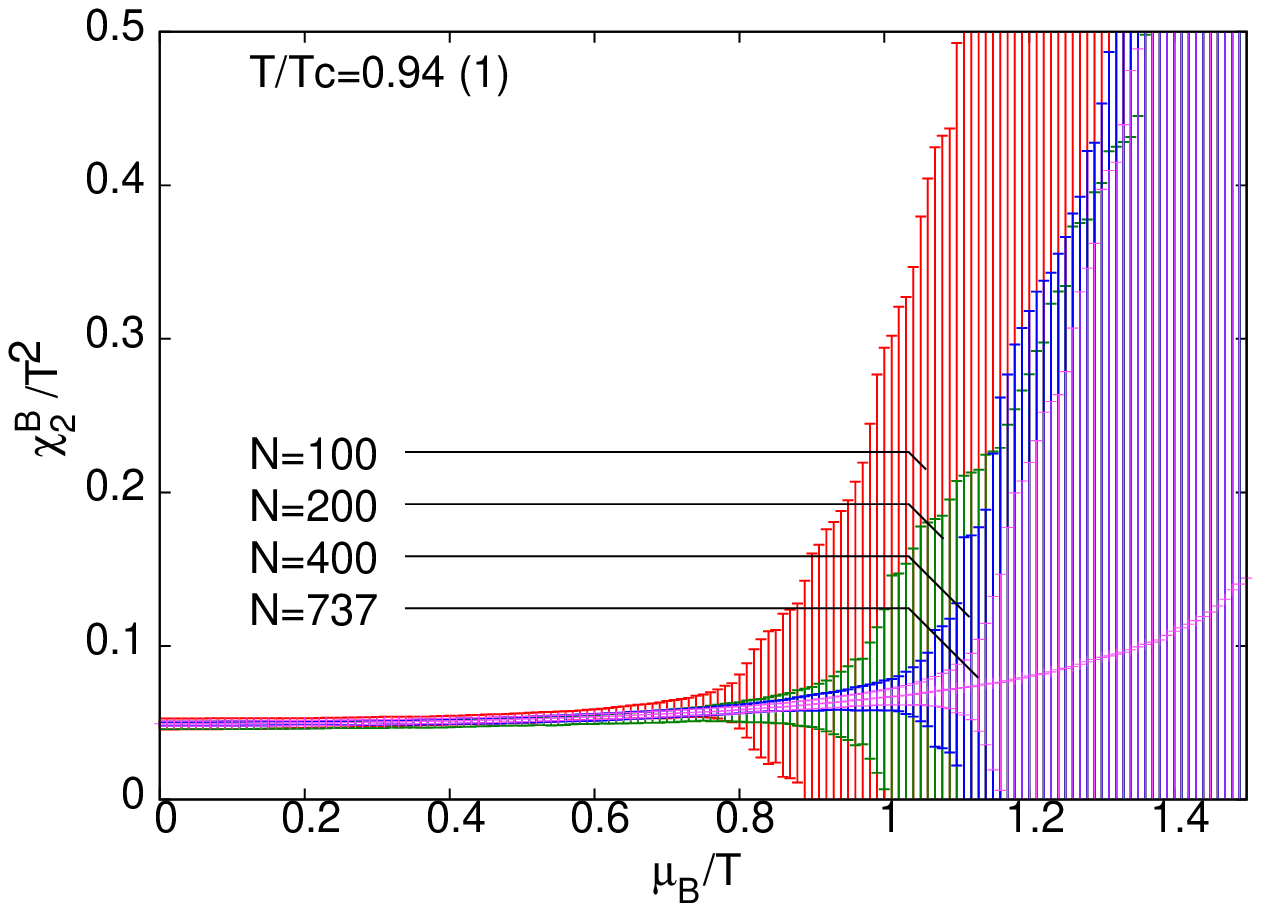}
\end{center}
\caption{Critical slowing down manifests itself as the increase in relative
 error as one extrapolates towards the critical point. As the statistics
 is increased, this blow-up of the error is postponed, but not removed.}
\eef{slow}

\bef
\begin{center}
\includegraphics[scale=1.0]{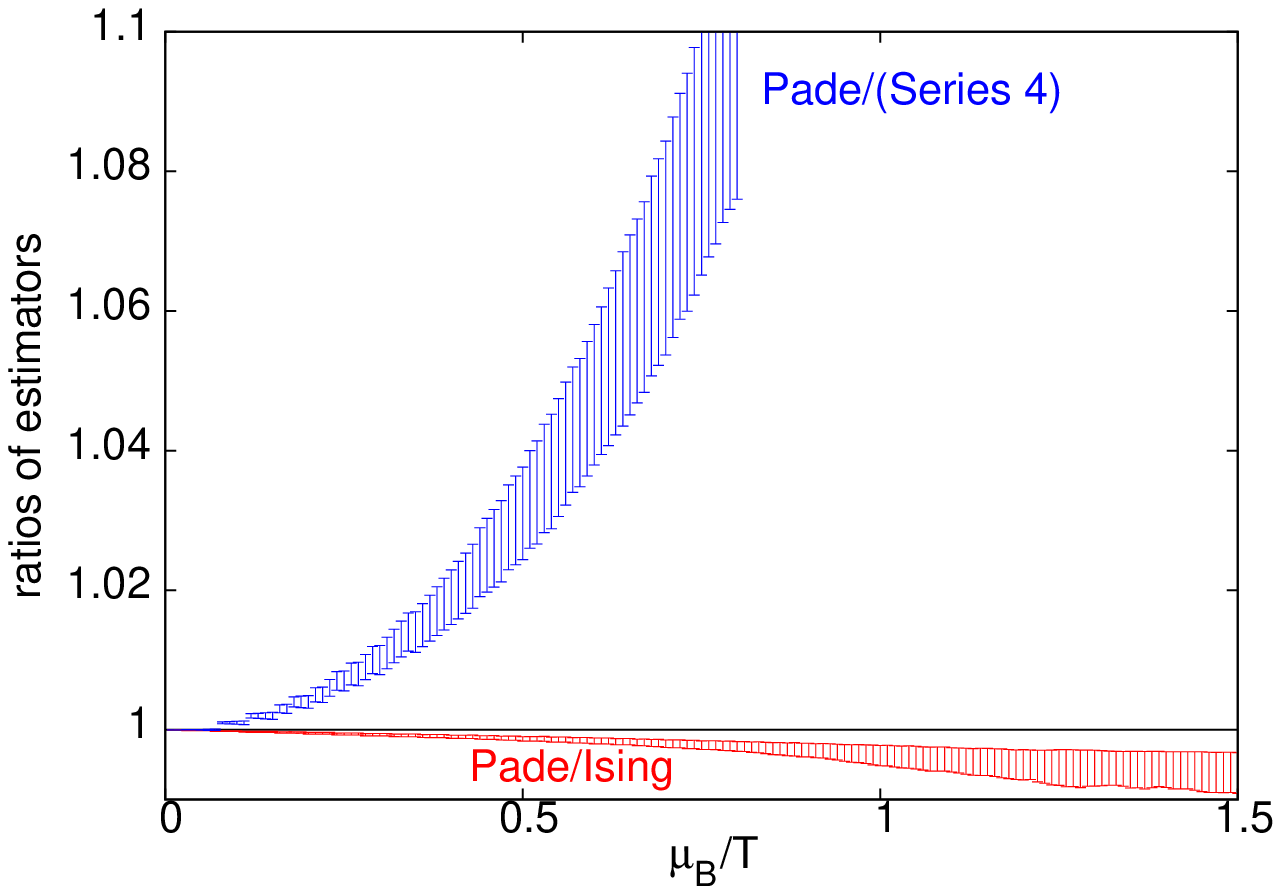}
\end{center}
\caption{Comparison of different estimators of $\chi_B^2(z,T)/T^2$ at
 $T=T_E$.  The ratio of the Pad\'e approximant to the series approximation
 which keeps all BNS up to order 4, called Series 4 here, is clearly
 different from unity, showing that the Pad\'e approximant contains more
 than two terms of the series. Also shown is the ratio of the Pad\'e
 approximant to the one with the Ising critical index; this is much
 closer to unity.}
\eef{approx}

\bef
\begin{center}
\includegraphics[scale=1.0]{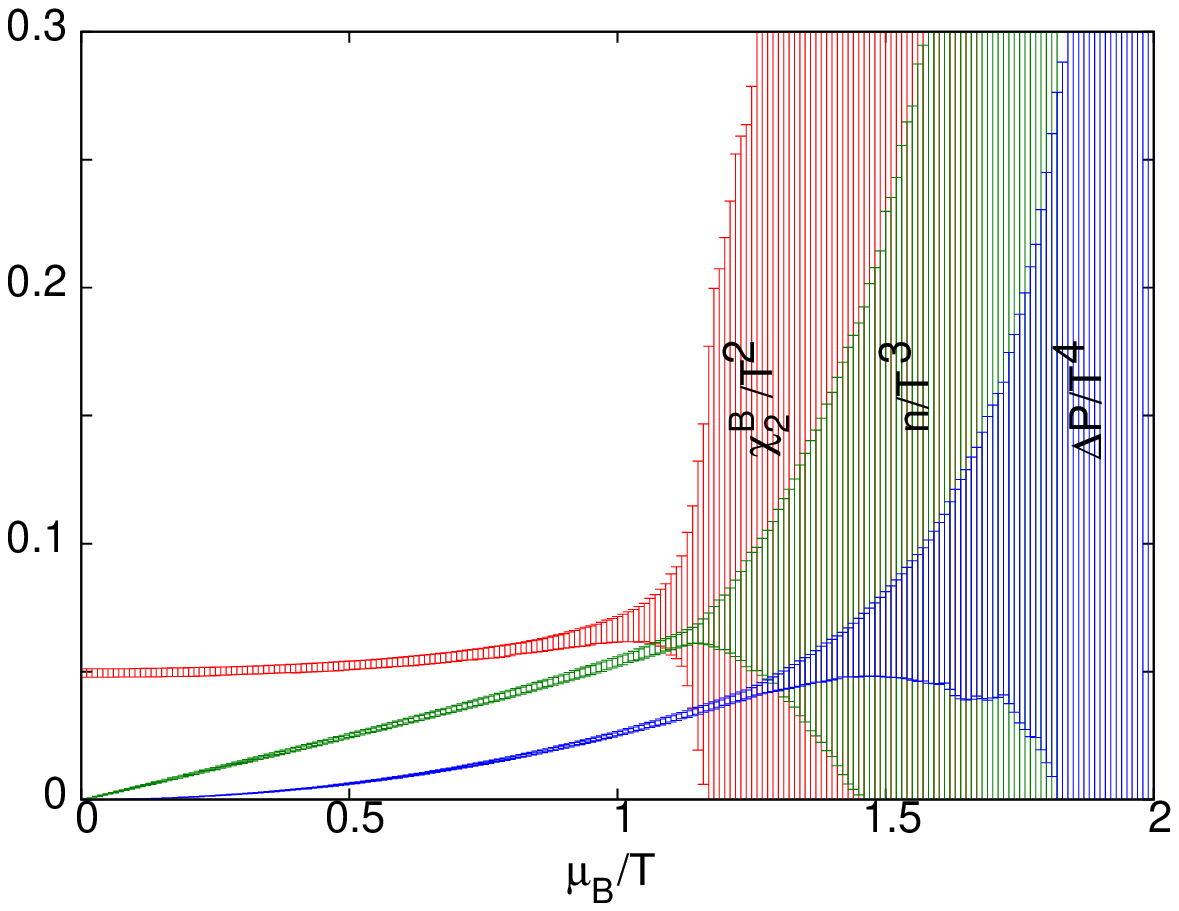}
\end{center}
\caption{$\chi_B/T^2$, $n/T^3$ and $\Delta P/T^4$ for $T/T_c=0.94\; (1)$
 obtained by successive integration. For the range of $z=\mub/T$ over
 which $\chi_B/T^2$ is constant, the baryon density is linear and the
 pressure is quadratic. The errors decrease with increasing order of
 integration because they are not pointwise errors, but induced through
 the errors in the Pad\'e parameters extracted for $m_1$.}
\eef{multiint}

\bef
\begin{center}
\includegraphics[scale=0.7]{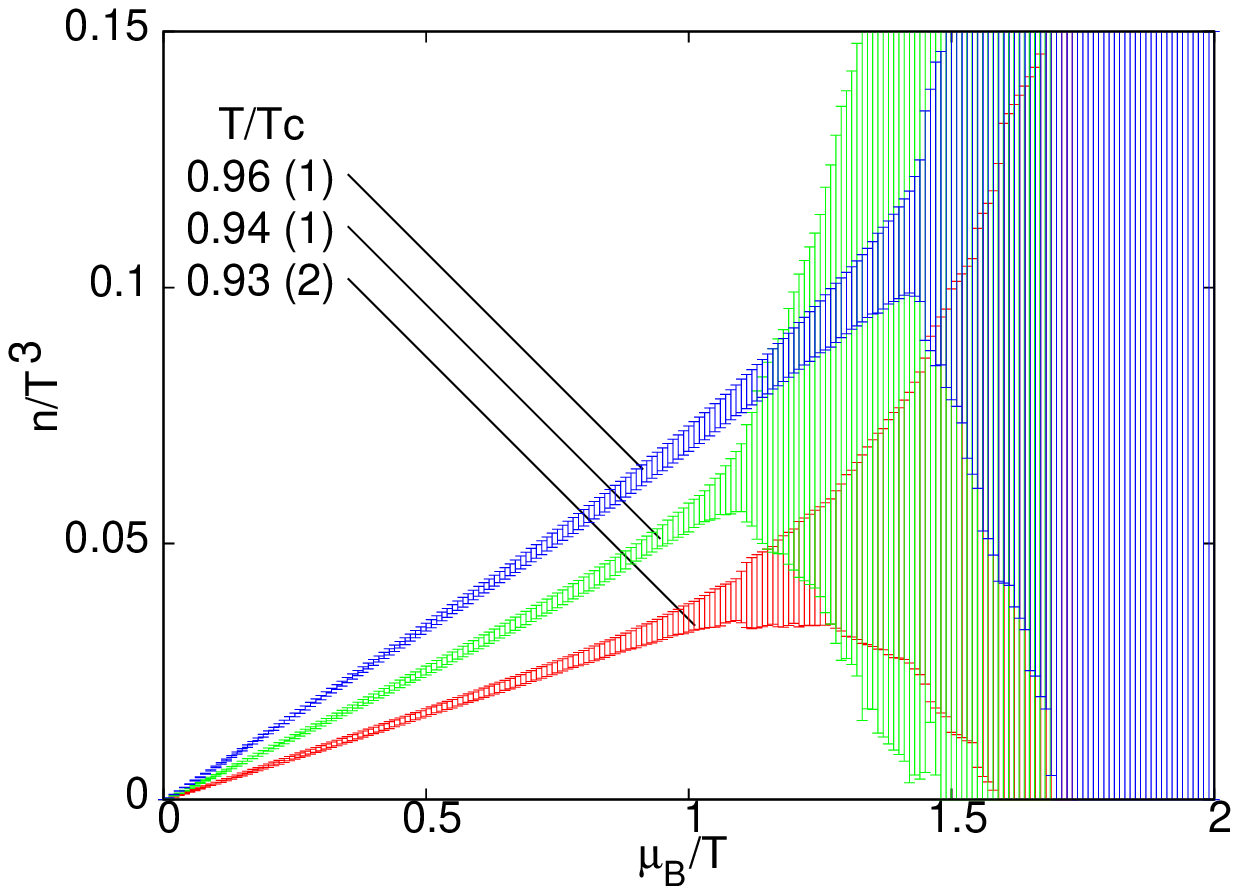}
\includegraphics[scale=0.7]{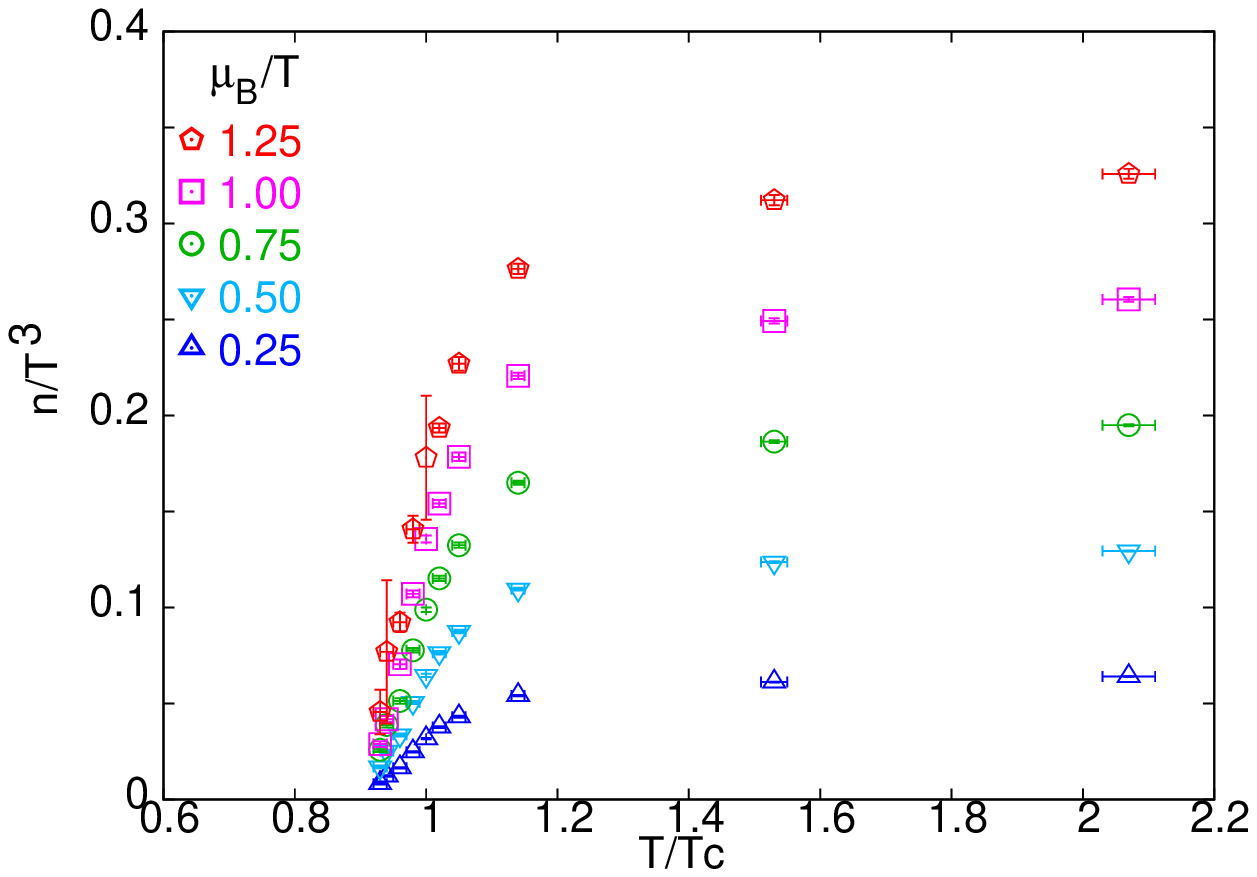}
\includegraphics[scale=0.7]{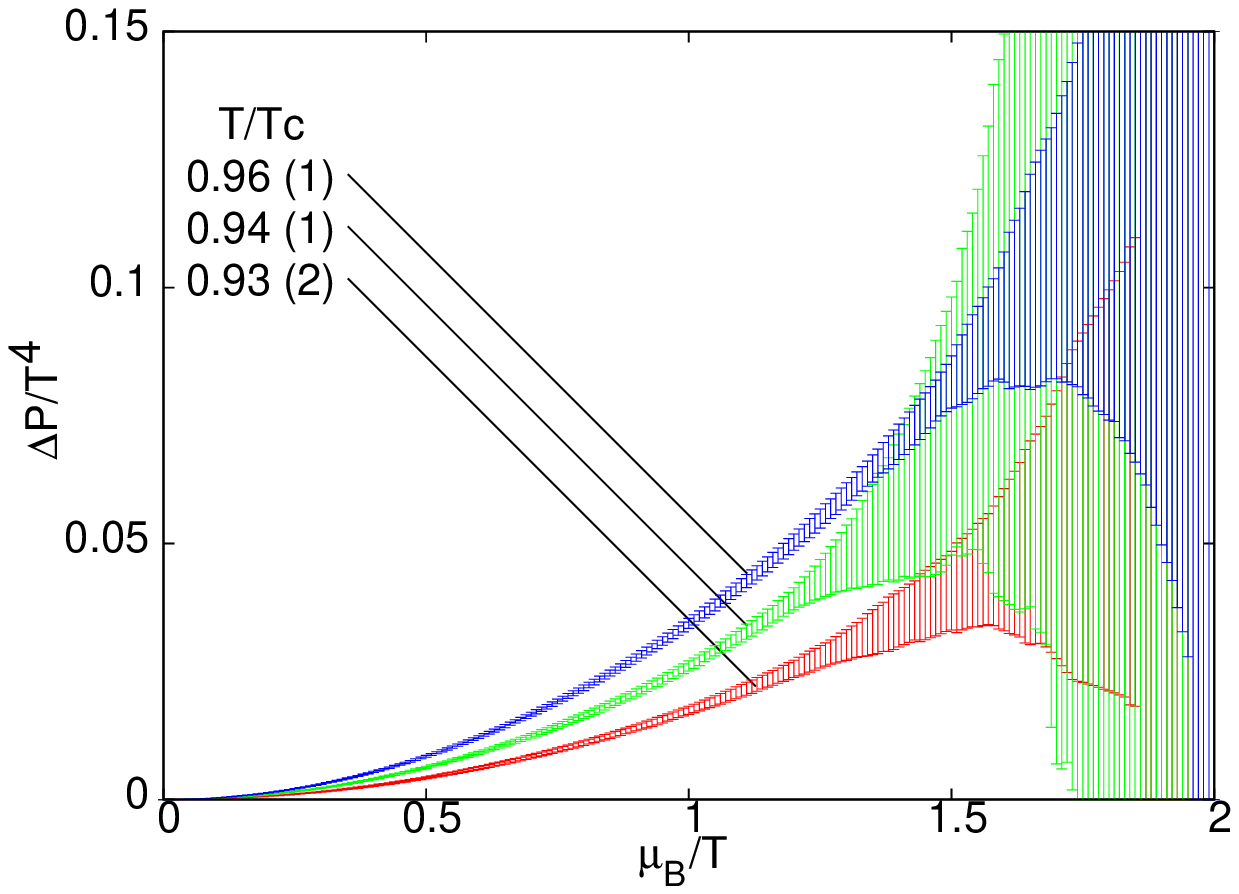}
\includegraphics[scale=0.7]{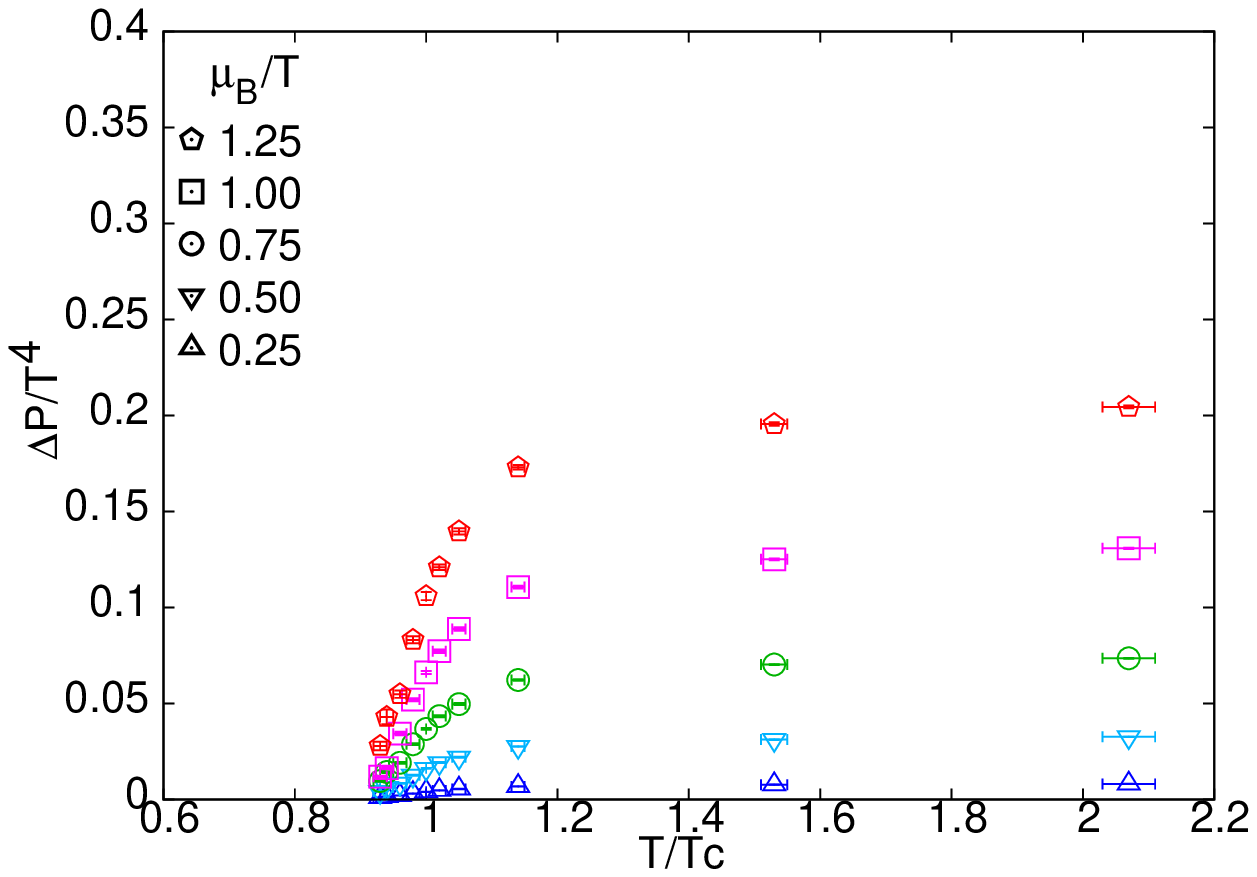}
\includegraphics[scale=0.7]{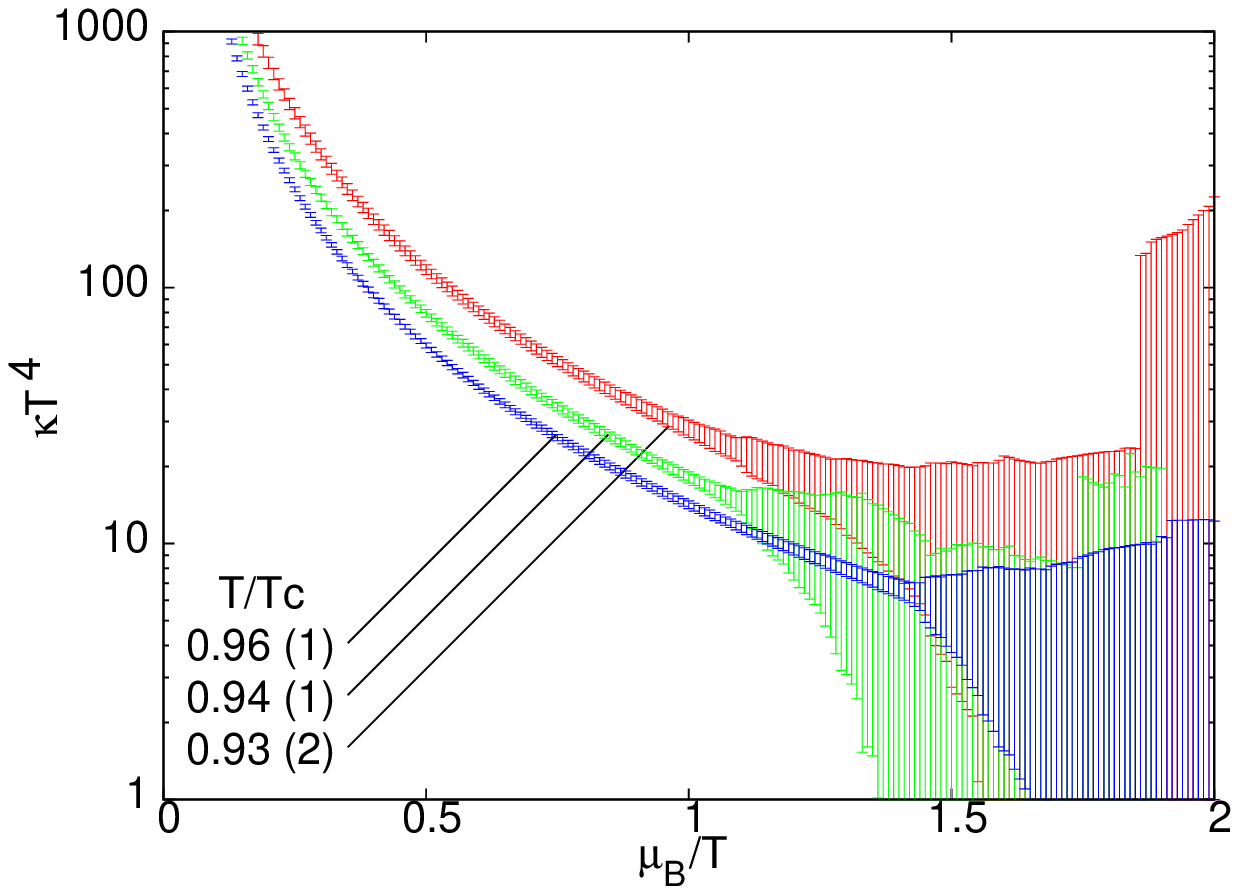}
\includegraphics[scale=0.7]{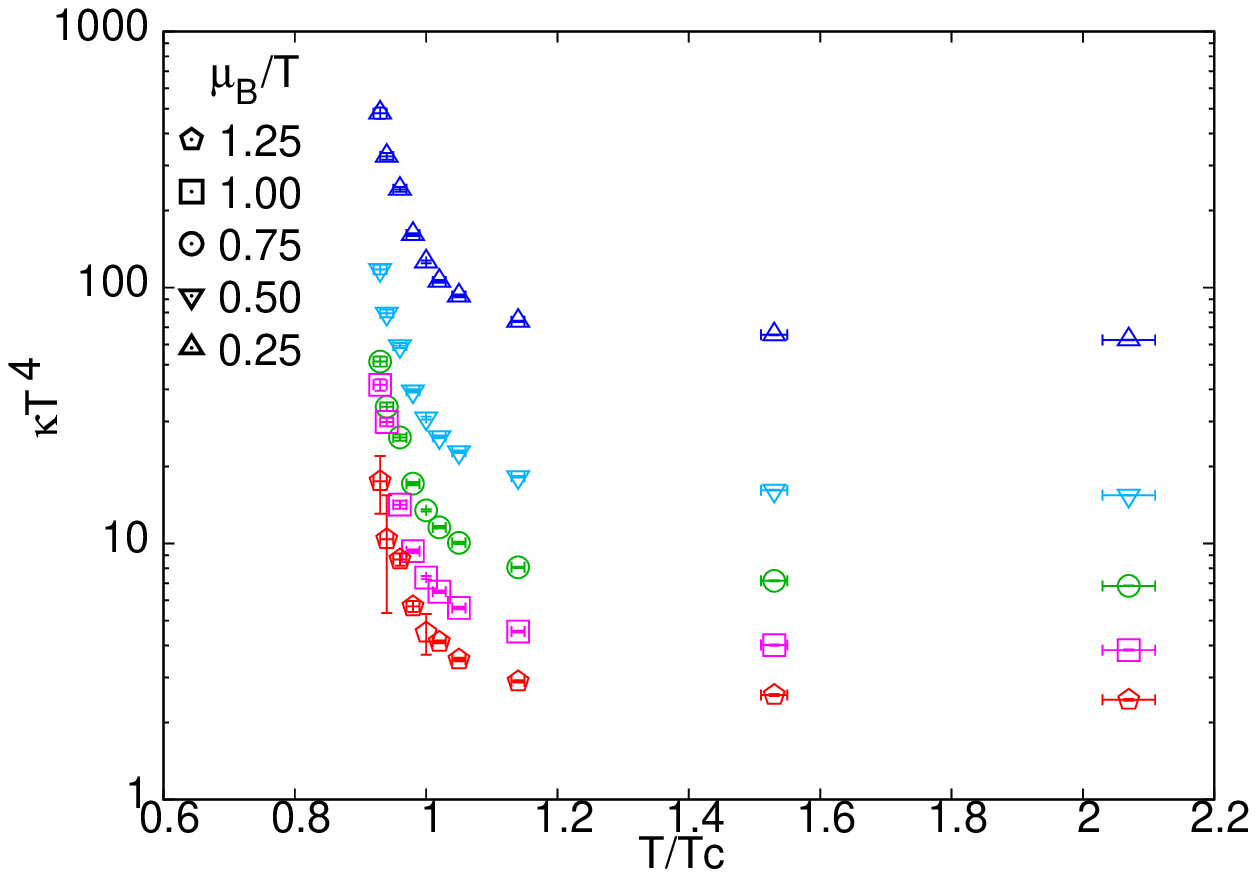}
\end{center}
\caption{The equation of state, $n(\mub,T)/T^3$ and $\Delta P(\mub,T)/T^4$,
 and the isothermal bulk compressibility, $T^4\kappa(\mub,T)$. These are
 insensitive to the lattice spacing for $T<T_c$.}
\eef{eos}

Having determined the BNS, \ie, the Maclaurin series coefficients of
$\Delta P$ and $n$, one can use these to determine the EOS at finite
$z$ through a truncated series expansion. This must be a reasonable
approximation to use when one needs the EOS at small $\mub$. However,
how small these $\mub$ should be is a question which must be determined
using the series itself. Clearly, in the temperature range close to our
estimated $T^E$ a truncated series expansion fails badly, because each of
the neglected terms could be as large as the terms included. In this case
one needs to resum the series. After this is done one can quantitatively
estimate the range of $\mub$ where the truncated series is useful.

A power divergence in $\chi_B$, as in \eqn{critd}, gives a pole in
\beq
   m_1(z,T) = \frac{\partial\log(\chi_B/T^2)}{\partial z}
    = \frac{\chi_B^3(z,T)/T}{\chi_B^2(z,T)/T^2}.
\eeq{m1def}
We can convert the series for $\chi_B/T^2$ into a series for $m_1$.
Since this has a simple pole, the series for $m_1/z$ can be approximated
by a $[0,1]$ Pad\'e approximant in $z^2$. The Pad\'e approximant for
$m_1$ is then of the form $2\psi z/(z_E^2-z^2)$. This two parameter form
resums the series in the whole region of $z$ where the singular form is
dominant \cite{pushan}. $\chi^2_B(z,T)$ is obtained by exponentiating the
integral of $m_1(z,T)$ found in this way.

The Pad\'e analysis yields the critical point, \ie, the position
of the pole, $z_E$, as well as the critical exponent $\psi$\footnote{With
better statistics we should be able to see the small difference between
the $z_E$ defined in this way and through \eqn{estmn}.}. The
current statistics is still not good enough to give a sharp estimate
of $\psi$.  An interesting point to note is that statistical errors in
the series coefficients translate into errors in the location of the
pole, thus leading to very large errors in $\chi_B$ in the vicinity of
$z_E$. With increasing statistics, this range of $z$ shrinks, as we show
in \fgn{slow}. The growth of the errors as one approaches the radius of
convergence is a manifestation of critical slowing down.

If one assumes that the critical exponent is given by the Ising
model, \ie, $\psi=0.79$ \cite{pushan}, then one has a one parameter
Pad\'e approximant to the series. This parameter is entirely fixed
by one term of the series.  The series expansion of the resulting
Pad\'e agrees within 68\% confidence limits with the next terms of
the known series.  In \fgn{approx} we show the ratio of the
Pad\'e and the Ising-constrained Pad\'e. They give similar results
for $\chi_B^2/T^2$. This implies, of course, that the Ising exponent
is compatible with this computation at the 90\% confidence level. This
computation is also consistent, at the same level, with the mean field
value $\psi=0.66$. In \fgn{approx} we also show the ratio of the
Pad\'e approximant and the series expansion up to $\chi^4_B$.
The ratio is very significantly different from unity, showing that the
higher order terms are clearly important.

The Pad\'e approximant is fitted using two terms of the series expansion
of $m_1$ (one if the Ising critical exponent is taken as an input). If
one re-expands the Pad\'e approximant in powers of $z$, then one has
predictions for the infinite series. We verified that the third term
of the series for $m_1$ (coming from the fourth term for $\chi_B$) is
consistent with the Pad\'e approximant re-expanded in this way. Similarly
we checked for consistency of two terms when the Ising critical exponents
is taken as an input.

Using $\chi^2_B$ we can make an estimate of the width of the critical
region. This is not a very well defined concept, but is usually taken as
an estimate of the region in which the regular part of the free energy is
negligible compared to the singular part. We convert this to a numerical
estimate by asking at which value of $z$ does $\chi_B$ become 5 times its
value at $T=T^E$ and $z=0$. By this criterion one enters the critical
region when $\mub=1.6 T^E$, \ie, at $\mub=0.87\mub^E$. If one asks the
more stringent question, when does $\chi_B$ become 10 times its value,
then we find that the critical region begins when $\mub=0.96\mub^E$. These
estimates use mean values. As one can see in \fgn{multiint}, the
uncertainties on these estimates are currently very large.

From this description of $\chi_B$ one obtains the EOS, namely
$n$ and $\Delta P$, by successive integrations. This is shown in
\fgn{multiint}. Notice that $\chi_B/T^2$ is nearly constant until
$z\simeq0.5$, as a result of which $n$ is close to linear and $\Delta P$
is almost quadratic. Note also that the range of critical slowing down
is smaller with increasing number of integrations. The reason for this
is simply that the errors shown in the figures are not pointwise errors
in $z$, but are induced by an error in $z_E$. With increasing number of
integrations, the singularity at $z_E$ becomes milder, as a result of
which the errors also become easier to control.

It is interesting that this analysis is also a good numerical description
of the data at $T/T_c\simeq2$. There the terms of the series expansion
beyond the fourth order are statistically insignificant. The Pad\'e
resummation deals with this by pushing the pole out to very large $z$.
At such temperatures we see that the truncated series and the Pad\'e
approximant gives similar results.

Our main results for the EOS are shown in \fgn{eos}. We display $n$,
$\Delta P$, and the isothermal bulk compressibility, $\kappa$, as
a function of $\mub/T$ at several fixed temperatures below $T_c$,
as well as a function of $T/T_c$ for several different values of
$\mub/T$. As before, there are two regimes of temperature: above and
below $T_c$.  In the region $T>T_c$, the EOS is dominated by the two
QNS $\chi_{20}$ and $\chi_{40}$ which are non-vanishing even in the
ideal gas. These have strong lattice spacing dependence, leading to
large cutoff dependence in the absolute value of the EOS in this high
temperature region\footnote{Usually improved fermion actions are used
to alleviate this problem. However, experience has shown that continuum
extrapolation remains important even so \cite{improv,baz}.  This 
improvement comes at a large cost in CPU time.}. However, a
comparison with the results of \cite{pushan} show that for $T<T_c$
the EOS is beginning to stabilize even for $\mub/T$ as large as 1.5.

\section{Fluctuations and Freezeout}\label{sec:m123}
\bef
\begin{center}
\includegraphics[scale=0.65]{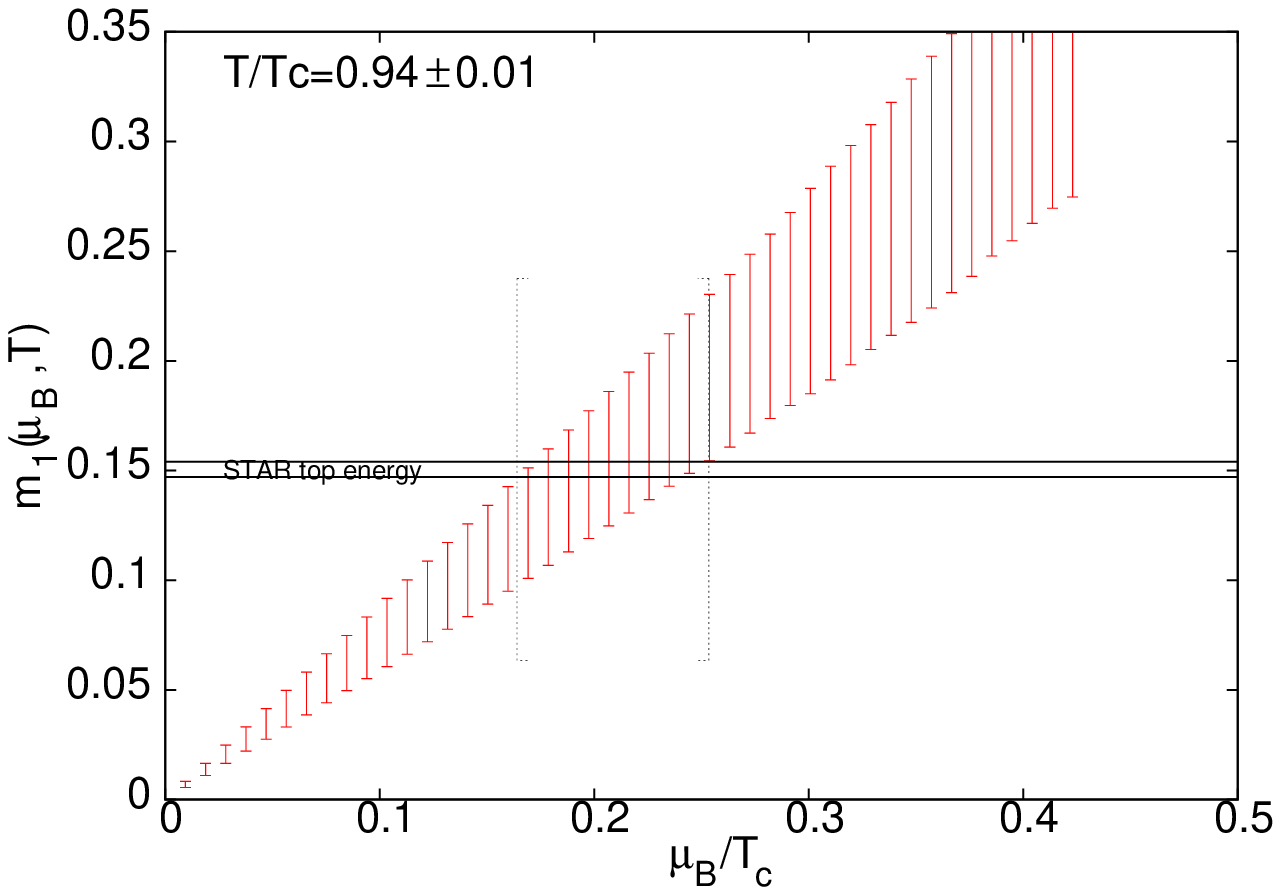}
\includegraphics[scale=0.65]{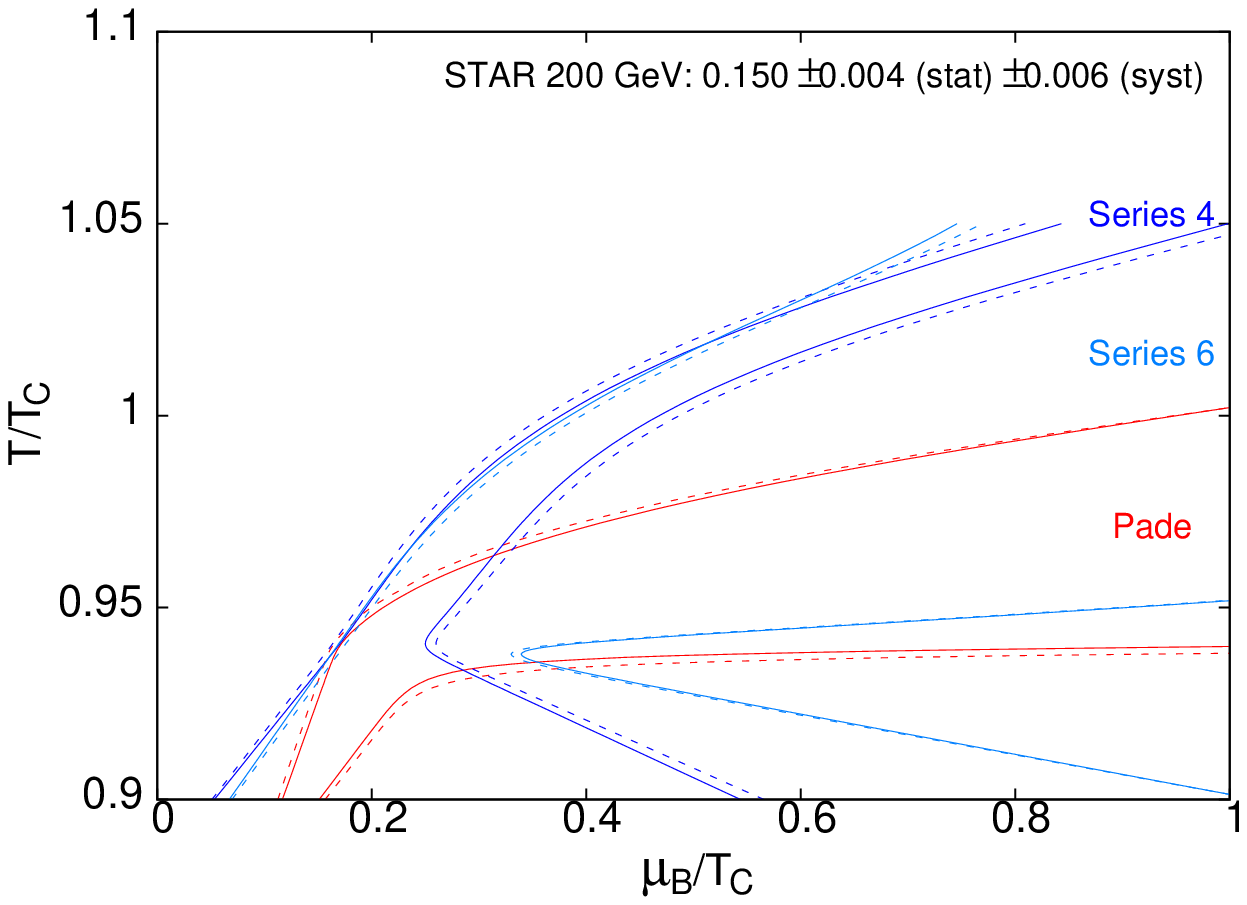}
\end{center}
\caption{In the first panel we show a comparison between our determination
 of $m_1^B$ and the STAR collaboration's measurement for this quantity in
 the fluctuation of protons at its highest energy. At each $T/T_c$, such
 a comparison yields a confidence interval for $z=\mub/T$ appropriate for
 the experimental result. In the second panel we show the smoothed freezeout
 band obtained by comparing lattice computations with the STAR data at the
 highest energy. The band labeled ``Series 4'' is obtained with the Taylor
 series including the fourth order BNS, ``Series 6'' up to sixth order,
 ``Pade'' for the Pad\'e approximant. The dashed line shows the effect of
 the experimental systematic error on the band.}
\eef{m1fix}

The study of QNS in lattice QCD has been interesting because of the
possibility of contact with experimental data. This may be confounded by
the fact that the fluctuations predicted by lattice computations are those
in the conserved baryon number, whereas those studied experimentally
are in the proton number. The connection between the two have been
addressed through hadronic Monte Carlos within event generators, and
seem to indicate that the comparison is safe. We accept this current
understanding in this section, while cautioning the reader that this might
change as the experimental error bars improve, or the inputs to the event
generators are updated.

Measures of fluctuations which have been discussed before \cite{cpod,plb} are
\beq
   m_1^B(\mub,T) = \frac{\chi^3_B(\mub,T)/T}{\chi^2_B(\mub,T)/T^2}, \quad{\rm and}\quad
   m_2^B(\mub,T) = \frac{\chi^4_B(\mub,T)}{\chi^2_B(\mub,T)/T^2},
\eeq{m123}
and their ratio $m_3^B =(\chi^3_B/T)/\chi_B^4$. The quantity
\beq
   m_0^B = \frac{n(\mub,T)}{\chi^2_B(\mub,T)}
\eeq{m0}
has also been proposed, as have various other combinations of these
ratios \cite{athanasiou}.  Such ratios can also be defined for charge
fluctuations.  The ansatz of \eqn{critd} has been used to investigate
the ratios in \eqn{m123}. Predictions for these observables in colliders
have also been made using the track of the freezeout points across
the phase diagram as one changes $\sqrt S$, the center of mass energy
of the collider \cite{plb}. First comparisons of lattice computations
with experimental results have successfully given rise to new ways of
approaching questions such as the physical value of $T_c$ \cite{glmrx},
or conversely, the freezeout conditions \cite{lfo}.

We have discussed the computation of $m_1^B$ in detail in the previous
section.  Here we follow a program outlined in \cite{cpod,plb} to
discuss a detailed comparison with data from \cite{starb}, in order to
find the point on the phase diagram where the fluctuations freeze out.
The extrapolation of $m_1$ in $\mub/T$ for each temperature can be
compared with the data to find possible range of chemical potentials
allowed by the data. We show an example in the first panel of \fgn{m1fix}
where the Pad\'e approximant to $m_1^B$ obtained at $T/T_c=0.94\pm0.01$
is compared to the data from the STAR experiment at $\sqrt S=200$ GeV
\cite{starb}. One can see that the error in the freezeout chemical
potential is dominated by the statistical error in the lattice
computation. The analysis shown in this panel of the figure is naive,
since it compares the 68\% probability bands of the experiment with
the lattice computation. We improve this estimate by putting such a
comparison within a bootstrap and extracting the 68\% confidence limits
on the fit using the bootstrap distribution of this estimate. We find
that this shifts the band marginally compared to the naive estimate. In
the same way, truncated series for $m_1$ can be used to determine freeze
out parameters.

One datum cannot determine both the parameters $\mub$ and $T$, so this
comparison gives us a strip of allowed values in the phase diagram,
as shown in the second panel of \fgn{m1fix}. We have used Bernstein
polynomials for smoothing. The figure shows differences between a
truncated power series and the Pad\'e approximant. Also the truncated
power series taken to different orders yields bands which are somewhat
different. Previous computations \cite{lfo} have used a truncated
power series taken to fourth order. We argued before that this misses
potentially large contributions from higher order computations. In the
figure we also show the contribution when the power series includes
the sixth order BNS, and when the Pad\'e resummation of the series
is performed. In the difference between the fourth and sixth order
series expansions, the systematic errors from the truncation of the
series are mixed with statistical errors in the determination of
the coefficients. The Pad\'e approximant is an attempt to remove the
systematic errors due to series truncation.  We expect that further
improvements in computation will shrink the error bands.

The further condition that freezeout occurs at the chiral cross over is
imposed in \cite{lfo}.  At this temperature\footnote{Due to the difference
in conventions used to define $T_c$, in our convention this implies
that $T/T_c\simeq0.94$.}, the truncated series method gives a freezeout
value for $\mub$ which roughly matches that found from a phenomenological
analysis of hadron yields \cite{hrg}. However, our analysis shows that
there are significantly larger theory uncertainties. Improvements in the
statistical errors in lattice computations will have many consequences
for the late-stage physics of heavy-ion collisions, including shedding
light on the very mechanism of freezeout.

\section{Conclusions}\label{sec:conclude}
In this paper we reported measurements of the QNS in QCD at finite
temperature with two flavours of light staggered quarks using lattices
with temporal extents of $N_t=8$. We have determined the $\mub=0$
cross over coupling $\beta_c=5.53$ using the peak in the Polyakov
loop susceptibility.  In \scn{method} we presented the evidence that
different numbers of source vector are required to obtain reliable
measurements of QNS at different orders. For the second order QNS a
hundred vectors is sufficient, but at least 1000 fermion source vectors
are needed to get good measurements for QNS of order 6 for $T<T_c$ whereas
substantially smaller number of source vectors are required in the high
temperature phase (see \fgn{syst}). We have used 2000 source vectors at
each temperature below $T_c$ and 800 or 1600 source vectors above $T_c$
in this study.

We compared measurements of the off-diagonal QNS $\chi_{11}/T^2$ made
with different lattice spacings in \fgn{ntdep}. This comparison indicates
that the $N_t=8$ results for this QNS may be close to the continuum
limit. The diagonal QNS $\chi_{20}/T^2$ is also shown in this figure.
Above $T_c$ there is clear evidence of finite lattice spacing effects
in $\chi_{20}/T^2$.  The reason for this is interesting. A particular
operator which contributes only to $\chi_{20}/T^2$ has large lattice
spacing effects for free staggered quarks. All other QNS are close to
the continuum limit predicted by weak-coupling theory, as shown in
\fgn{vecdephi}.  In the temperature range below $T_c$ there is good
agreement between earlier results for $\chi_{20}/T^2$ for $N_t=6$
\cite{nt6} and the new results. This indicates that in the confined
state these results may be close to the continuum limit.

We have also shown our measurements of several of the higher order
QNS in \fgn{series}. There is structure visible in these QNS in the
neighbourhood of $T_c$.  The fact that these measurements are made
using the same configurations and source vectors makes them strongly
correlated with each other, something that the error bars shown in the
figures cannot capture. This has consequences for all derived measurements.

At $T/T_c=0.94\pm0.01$ we find the radius of convergence of the Taylor series
expansion for the BNS is $\mu/T=1.85\pm0.04$ (see \fgn{endpoint}). At this
temperature all the terms in the series which we can measure turn out to
be positive, implying that the singular point is on the real axis. This
leads us to believe that the radius of convergence identifies a critical
point of QCD. We note that this estimate is completely consistent with
that presented earlier for $N_t=6$ in \cite{nt6}.

The existence of a finite radius of convergence of a series expansion
is a statement of the mathematical fact that the successive terms
of the series become equal at the radius of convergence. This means
that in deriving consequences from the series, one cannot afford to
truncate the series, but must attempt to sum all the terms. A method of
doing this through Pad\'e approximants was first used in \cite{pushan}
with lattice spacing which is about twice of what we use here. This
also gives an independent, though coarse, check on the estimate of
the radius of convergence referred to earlier. We have used the same
method in \scn{eos} to continue the equation of state to finite chemical
potential. Our results for the baryon number density, $n(\mub,T)/T^3$,
as a function of temperature and chemical potential, the change in the
pressure due to the chemical potential, $\Delta P(\mub,T)/T^4$, and the
isothermal bulk compressibility, $T^4\kappa$, are shown in \fgn{eos}.

It is known that direct Monte Carlo simulations suffer from critical
slowing down, due to the unbounded growth of correlations near a critical
point. A similar phenomenon is observed in the reconstruction of physical
quantities using series expansions, as demonstrated in \fgn{slow}. In
spite of this, we are able to extract the equation of state up to
$\mub/T\simeq1.25$ with reasonable accuracy.

In \scn{m123} we examine the influence of the full series expansion in
connecting to data on fluctuations. We show that truncating the series
expansion \cite{lfo} leads to changes in the determination of the
freeze out conditions for fluctuations. This includes statistical and
systematic theory uncertainties. We have shown, in the right hand panel of
\fgn{m1fix}, that the results shift when truncating the series expansion
up to the fourth order BNS or the sixth order. We have also shown the
results obtained when we try to estimate the complete series by a Pad\'e
resummation.  These uncertainties can be bounded better in future by going
to larger orders in the series expansion.  The full analysis which was
suggested in \cite{plb} is needed to make contact with experimental data.
This is, of course, equally true for the equation of state.

We have examined a previously under-appreciated source of systematic
errors in the reconstruction of physical quantities from Taylor series
expansions, namely the truncation errors in the series expansion. While
our measurements are restricted to moderate lattice spacings, \ie,
$1/(8T)$, we have presented some evidence that the systematic errors
arising from extrapolations in lattice spacing are much smaller than these
truncation errors. Future work will concentrate on reducing statistical
and both kinds of systematic errors.

{\bf Acknowledgments}: The computations described here were carried
out on the IBM BlueGene P of the ILGTI in TIFR. We thank Ajay Salve and
Kapil Ghaliadi for technical support.  We would like to thank Alexi
Vuorinen and Sylvain Mogliacci for sharing code and tables of the DR
computations.

\end{document}